\documentclass[aps,prd,showpacs,nofootinbib,preprintnumbers]{revtex4}
\usepackage{graphicx,graphics}
\usepackage{amssymb}
\usepackage{amsmath}
\usepackage{pifont}
\usepackage{natbib}
\usepackage{subfigure,epsfig,epstopdf}
\usepackage{color}
\usepackage{bm}

\newcommand{\be}{\begin{equation}}
\newcommand{\ee}{\end{equation}}
\newcommand{\bea}{\begin{eqnarray}}
\newcommand{\eea}{\end{eqnarray}}

\def\siml{{\ \lower-1.2pt\vbox{\hbox{\rlap{$<$}\lower6pt\vbox{\hbox{$\sim$}}}}\ }}

\begin{document}
\title{Heavy Quarkonium moving in a Quark-Gluon Plasma}
\author{Miguel Angel Escobedo}
\affiliation{Physik-Department, Technische Universit\"at M\"unchen, James-Franck-Str. 1, 85748 Garching,
Germany
}
\author{Floriana Giannuzzi}
\affiliation{Universit\`{a} degli Studi di Bari and Istituto Nazionale di Fisica Nucleare, Sezione di Bari,
                          via Orabona 4, I-70126, Bari, Italy
}
\author{Massimo Mannarelli}
\affiliation{
I.N.F.N., Laboratori Nazionali del Gran Sasso, Assergi (AQ), Italy}
\author{Joan Soto}
\affiliation{Departament d'Estructura i Constituents de la Mat\`eria and Institut de Ci\`encies del Cosmos\footnote{Associated with the Institut de Ci\`encies de l'Espai (CSIC)}, Universitat
de Barcelona\\
Mart\'\i$\;$ i Franqu\`es 1, 08028 Barcelona, Catalonia}
\preprint{UB-ECM-PF-13/88}
\preprint{ICCUB-13-064}
\preprint{BARI-TH/13-674}
\preprint{TUM-EFT35/12}
\pacs{11.10.St, 14.40.Pq, 32.70.Cs, 36.10.Ee}

\begin{abstract}
By means of effective field theory techniques, we study the modifications of some properties of weakly coupled heavy quarkonium states propagating through a  quark-gluon plasma at temperatures much smaller than the heavy quark mass, $m_Q$. Two different cases are considered,  corresponding to two different hierarchies between the typical size of the bound state, $r$, the binding energy, $E$, the temperature, $T$, and the screening mass, $m_D$. The first case corresponds to the hierarchy $m_Q \gg 1/r \gg T \gg E \gg m_D$,  relevant for moderate temperatures,
and the second one to the hierarchy   $m_Q \gg T \gg 1/r \,, m_D \gg E$, relevant for studying the dissociation mechanism. In the first case we determine the perturbative correction to  the binding energy and to the decay width of states with arbitrary angular momentum, finding that the width  is a decreasing function of the velocity. A different behavior characterizes the second kinematical  case, where the width of $s$-wave states    becomes a non-monotonic function of the velocity, increasing at moderate velocities and decreasing in the ultra-relativistic limit. We obtain a simple analytical expression of the decay width for $ T \gg 1/r \gg m_D \gg E$ at moderate velocities, and we derive the $s$-wave spectral function for the more general case $T \gg 1/r \,, m_D \gg E$.
A brief discussion of the possible experimental signatures   as well as a comparison with the relevant lattice data are also presented.
\end{abstract}

\maketitle

\section{Introduction}
Heavy quarks produced in the early stage of  relativistic heavy-ion collisions  are valuable probes of the medium that develops at later stages. They can be used to resolve its energy density, and eventually to understand which are the fundamental degrees of freedom of the system. Indeed, at sufficiently high energy densities matter should form  a quark-gluon plasma (QGP) and the propagating heavy quarks should be capable to convey this information to us. In relativistic heavy-ion collisions  (HIC)  it is expected that  the variation of the interaction between  heavy quarks due to the creation of a  hot medium should be observable.
 In particular, the Debye screening of the Coulomb-like potential between a heavy quark and a heavy antiquark was proposed in Ref.~\cite{Matsui:1986dk, Karsch:1987pv} as a dissociation mechanism, resulting in the suppression of the yields of heavy quarkonium, $Q \overline Q$,   states in HIC.  The low-lying heavy 
quarkonium states are considered as the most powerful probes because they are the only hadronic states that are able to survive above the deconfinement temperature (see \cite{Rapp:2008tf,Mocsy:2013syh} for reviews). This is due to the fact that even at weak coupling, namely ignoring confinement, these states still exist. In addition, the vector states enjoy a rather clean dilepton decay channel, which makes them easy to identify experimentally, although in HIC 
the corresponding   background  is not yet completely understood, see \textit{e.g.} \cite{Abelev:2010am}. 
  
Suppression of charmonium states was first observed in Pb-Pb collisions by the NA50 Collaboration~\cite{Abreu:1997jh} at relatively low center of mass energy (per nucleon), $\sqrt{S_{NN}} = 158 $ GeV. However,
 in contrast with the naive Debye screening scenario, a complicated pattern  emerged because of the various processes involving charm quarks, see \textit{e.g.} \cite{Adare:2011yf} for a recent experimental analysis. Charm quarks can indeed be produced not only by hard scattering at the early stage of the collision (prompt production), but also later on by  collisions inside the QGP (non-prompt production) and their density is sufficiently high that they can recombine in charmonium states.   Bottomonium states give a much clearer signal,   see  \textit{e.g.} the discussion in \cite{Strickland:2012as}, because these states are very massive and can only be promptly produced  by hard scatterings: the probability of generating  these states within the QGP is extremely low. Moreover, since bottomonia are heavy and  compact objects,  they do not equilibrate with the expanding medium,  and can be really considered as external probes. 

Recent  experimental results by the CMS Collaboration \cite{Silvestre:2011ei, Chatrchyan:2011pe,Chatrchyan:2012fr} and by the STAR Collaboration \cite{Reed:2011fr}  indicate a clear  suppression of $\Upsilon$ states, meaning that the nuclear modification factor  $R_{AA}$, expressing the ratio of the yields of  a state in HIC  with respect to $p\,$-$p$ collisions, is less than $1$. The $R_{AA}$ decreases with increasing centrality and/or with increasing $p_T$ and is higher  for the fundamental state, meaning that it is less suppressed.   Indeed, the experimental results of   Pb\,-Pb collisions at $\sqrt{S_{NN}} = 2.76 $ TeV \cite{Chatrchyan:2012fr} seem to indicate sequential suppression of bottomonium states \cite{Chatrchyan:2011pe, Reed:2011fr, Chatrchyan:2012fr}, in particular, integrating over centrality it has been found that      $R_{AA} (\Upsilon (1s)) \simeq 0.6$, $R_{AA} (\Upsilon (2s)) \simeq 0.1$ and $R_{AA} (\Upsilon (3s)) < 0.1$.

The effective field theory (EFT) techniques are very useful for the description of heavy quarkonia, because they are suited for handling systems with well separated  energy scales. In the case of $Q \overline Q$ in a thermal medium ~\cite{Escobedo:2008sy,Brambilla:2008cx,Brambilla:2010vq,Escobedo:2010tu}  two distinct kinds of scales appear, namely the non-relativistic scales and the thermal scales. The  non-relativistic scales are given by the mass of the heavy quark, $m_Q$, the typical momentum transfer, $1/r \propto  m_Q \alpha_s$ ($\alpha_s=g^2/4\pi$ is the QCD coupling constant), and the binding energy, $E \propto  m_Q \alpha_s^2$ \cite{Caswell:1985ui}. We have assumed the weak coupling regime and identified the relative velocity between $Q$ and $\bar Q$ with $\alpha_s$ (see \cite{Brambilla:2004jw,Pineda:2011dg} for reviews). The relevant thermal scales to our analysis are the temperature $T$ and the Debye mass $m_D \propto g T$.
We shall discuss two possible hierarchies $m_Q \gg 1/r \gg T \gg E \gg m_D$ and $m_Q \gg T \gg 1/r \,, m_D   \gg E$, which we shall refer to as Case I and Case II, respectively.  If the bound state moves with respect to the medium, the EFT analysis becomes  more complicated, because additional energy scales may appear \cite{Escobedo:2011ie}. We shall restrict our analysis here to the case of moderate velocities ($v\nsim 1$) for which no further scales are induced, so that Case I and Case II above can be safely addressed. However, at some instances we will push our results to the ultra-relativistic limit ($v \rightarrow 1$). Although this gives the correct results in the QED case \cite{Escobedo:2011ie}, and hence, we expect them to be sensible for QCD as well, one must keep in mind that they are on a less firm ground.

Using EFT techniques it has been shown 
that, at least in perturbation theory, the dissociation of heavy quarkonia is not due to the Debye screening but to the appearance of an imaginary part 
in the potential~\cite{Laine:2006ns,Escobedo:2008sy,Brambilla:2008cx,Laine:2008cf}. In other words, at high temperature heavy quarkonia disappear not because the binding energy vanishes, but because the \textit{thermal width} becomes so large that the $Q \overline Q$ state melts in the continuum. 
In QCD two different processes  contribute to the thermal width: 
inelastic parton scattering, which is the dominant process for  $m_D \gg E$, and the  gluo-dissociation process 
 that corresponds to the decay of a color singlet state into a color octet induced by a thermal gluon;  this process is dominant for $m_D \ll E$ \cite{Brambilla:2011sg,Brambilla:2013dpa} (see \cite{Grandchamp:2001pf} for an early discussion). The inelastic parton scattering is often referred in the literature as Landau damping, the reason is that this scattering is always mediated by a space-like gluon and can be related to 
the absorptive part of the gluon propagator. We shall also use this nomenclature from now on.
In the strong coupling regime, the effect of an imaginary part in usual potential models has been addresed in \cite{Petreczky:2010tk}, and in the so called T-matrix approach imaginary parts are incorporated in heavy quark self-energies through a set of Schwinger-Dyson equations, see for instance \cite{Mannarelli:2005pz,Riek:2010fk}. Recently, the imaginary part of the potential has also been calculated on the lattice \cite{Rothkopf:2011db,Burnier:2012az} (see also \cite{Akamatsu:2011se,Akamatsu:2012vt}, for a description in terms of open quantum systems).

The study of bound states propagating in the QGP at  finite velocity  is relevant for $\Upsilon$ states that are  promptly produced in  HIC and   will cross the hot medium with a relative velocity $v$.  
In principle it might happen that heavy flavors are drifted by the expanding plasma. Indeed the 
 PHENIX Collaboration~\cite{Adare:2006nq, Awes:2008qi}   has observed a large $v_2$ of heavy-flavor electrons, suggesting that there is significant damping of heavy quarks while they travel across the medium. This picture has also received support from microscopic calculations of heavy quark diffusion in the quark-gluon plasma~\cite{vanHees:2007me}.
However, the elliptic flow of the   $\Upsilon$ induced by the expanding medium should be negligible if the Debye length is larger than the typical distance between quarks, because heavy quarkonium at distances larger than its radius is colorless.  Therefore, in both Case I and Case II the drift should be small 
and  certainly less important for bottomonia than for lighter quarkonia,  like the $J/\psi$, which can be non-promptly produced and are expected to  roughly comove with the thermal bath. This is because before recombining both charm quarks have been drifted by the expanding QGP. 

In the first study of moving $Q \overline Q$s   performed in~\cite{Chu:1988wh},  the hierarchy of scale of Case II was assumed, but only the real part of the potential was considered. The imaginary part of the potential was studied in QED in \cite{Escobedo:2011ie}, where  the velocity dependence of the cylindrically symmetric real and imaginary parts of the potential were determined. In the present paper we extend the analysis of  \cite{Escobedo:2011ie} in two directions. 

Regarding Case I, we consider QCD instead of QED; the main difference is that while in QED a proton and an electron will always form an electrically neutral state, in QCD a heavy quark and a heavy antiquark can be found in a singlet and an octet state, and this induces new terms in the computation. We determine the velocity dependence of the thermal width and of the energy shifts at the leading order. In particular, we find that at the leading order the energy shifts of the $s$-wave states do not depend on the velocity (like in QED), but the energy shifts of all the other states depend on the velocity (unlike in QED). 

Regarding Case II,  we extend the analysis of \cite{Escobedo:2011ie} by deriving an approximate analytical expression for the $s$-wave width as a function of the temperature and of $v$, valid for the particular hierarchy of scales $T\gg 1/r\gg m_D\gg E$. Moreover,  considering the more general case, where $T\gg 1/r$ and $m_D\gg E$ but the product   $r m_D$ is arbitrary, we solve the corresponding Schr\"odinger equation numerically and determine the spectral representation of the two-point function.

This paper is organized as follows. In Sec.~\ref{sec:CaseI} we discuss Case I, corresponding to the hierarchy  $m_Q \gg 1/r \gg T \gg E \gg m_D$. We derive the expression of the width and of the energy shifts as a function of the temperature and of the velocity of the bound state. In Sec.~\ref{sec:CaseII} we discuss Case II, corresponding to the hierarchy  $m_Q\gg T \gg 1/r  \sim m_D\gg E$, 
this section is divided in two subsections, in the first one we do an analytical analysis of the case $1/r\gg m_D$ while in Sec.~\ref{sec:CaseII_spectral} we solve the Schr\"odinger equation numerically for $1/r\sim m_D$ and determine the  spectral function for various values of $T$ and $v$. In Sec. \ref{sec:Conclusion} we present a brief discussion of the observable consequences of  the velocity dependent thermal width, we compare our results with existing lattice simulations and we draw our conclusions.  In the Appendix \ref{appendix-gen} we  discuss the framework  used  to take into account the effect of a moving thermal medium. In the Appendix \ref{appendix-osm} we present some details and numerical checks of the  procedure used in Sec.~\ref{sec:CaseII_spectral} to derive the spectral amplitudes.

\section{Case I
}\label{sec:CaseI}

The low-lying bottomonium states, $\Upsilon(1s)$ and $\eta_b$, produced at early times in relativistic heavy-ion collisions 
are likely to have a typical size, $r$,  smaller than the inverse temperature during most of their evolution in the QGP. At intermediate times, the temperature is also likely to be larger than the binding energy. 
Having in mind this possibility we shall study in detail the particular case
\be m_Q \gg 1/r\gg T\gg E\gg m_D\,.\label{eq:hierarchy1} \ee 
This hierarchy of energy scales was considered in \cite{Brambilla:2010vq} for a thermal bath at rest. For a moving thermal bath, it was studied in full detail for the  hydrogen atom in \cite{Escobedo:2011ie}; here we generalize those results to  QCD. 

The general formalism 
to deal with a moving thermal medium is reviewed in the Appendix \ref{appendix-gen}. An important outcome  is that
in the bound state reference frame two additional energy scales should be considered: 
\be T_+=T\sqrt{\frac{1+v}{1-v}} \qquad {\rm and} \qquad T_-=T\sqrt{\frac{1-v}{1+v}}\,,\label{Tpm}\ee 
where $v$ is the velocity of the medium with respect to the bound state. When $v \to 1$ these scales are widely separated, a fact that must be taken into account in order to build the appropriate effective field theory (EFT). For instance, in \cite{Escobedo:2011ie} the appropriate EFT was constructed for the case $T_+ \gg 1/r \gg T$, which is different from the EFT obtained for the case $T_+ \sim T \sim T_-$, valid for $v \nsim 1$. We shall mainly restrict ourselves to the latter case, and only comment on the limit $v \rightarrow$ 1. Note, indeed, that the QED analysis of \cite{Escobedo:2011ie} shows that the results obtained  with the EFT theory valid for $T_+ \sim T \sim T_-$  and then naively extrapolated to the $v \to 1$ case, coincide with the ones obtained with the proper EFT with $T_+ \gg 1/r \gg T$ (if no large log resummations are performed). Hence, our results may hold for the $v\to 1$ case as well.

\subsection{Matching between pNRQCD and pNRQCD$_{\rm HTL}$}
Since $1/r\gg T$ we can take as the starting point the pNRQCD Lagrangian at $T=0$ \cite{Pineda:1997bj,Brambilla:1999xf}, which is obtained from QCD by sequentially integrating out energy scales of order $m_Q$ and of order $1/r$,
\begin{eqnarray}
{\cal L}_{\textrm{pNRQCD}} &=& 
{\cal L}_{\rm light}
+ \int d^3{\bf r} \;\left(\!\!\!\!\!\phantom{\frac 1 r}{\rm Tr}\left\{ {\rm S}^\dagger \left[ i\partial_0 - h_s \right] {\rm S} 
+ {\rm O}^\dagger \left[ iD_0 -h_o  \right] {\rm O} \right\}\right.
\nonumber\\
&+& \left. {\rm Tr} \left\{  {\rm O}^\dagger \mathbf{r} \cdot g\mathbf{E} \,{\rm S}
+ {\rm S}^\dagger \mathbf{r} \cdot g\mathbf{E} \,{\rm O} \right\} 
+ \frac{1}{2} {\rm Tr} \left\{  {\rm O}^\dagger \mathbf{r}\cdot g\mathbf{E} \, {\rm O} 
+ {\rm O}^\dagger {\rm O} \mathbf{r} \cdot g\mathbf{E}  \right\}  + \dots \right)\,,
\label{pNRQCD}	
\end{eqnarray}
where ${\cal L}_{\rm light}$ is the QCD Lagrangian for light quarks, $g$ is the coupling constant, $\bm E$ is the chromo-electric field,  S and O are   the quark-antiquark singlet and octet fields respectively, and 
\be
h_{s,o} = \frac{\bm p^2}{m_Q} + 
V_{s,o}+\cdots \quad , \quad  V_s=-\frac{C_F\alpha_s}{r} \quad , \quad  V_o=\frac{(C_A/2-C_F)\alpha_s}{r}\,,
\ee   
($r=\vert {\bf r}\vert$) correspond to the singlet and octet Hamiltonians (the dots stand for $1/m_Q$ corrections); hereafter $C_A=3$ and $C_F=4/3$. Thermal corrections to this Lagrangian are exponentially suppressed
 because the energy scales integrated out ($m_Q$ and $1/r$) are much larger than $T$. Note also that no dependence on the velocity appears at this stage because the velocity enters in the calculation through the scales $T_+$ and $T_-$ in (\ref{Tpm}) only.

\begin{figure}
\includegraphics[scale=1]{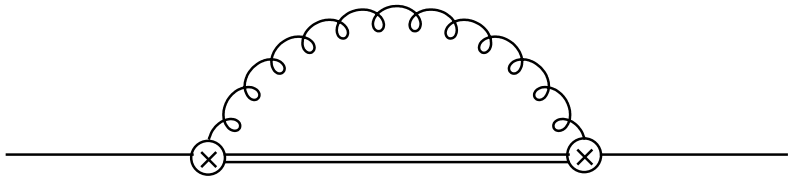}
\caption{The singlet self-energy. The plain line represents the singlet field, the double line represents the octet and the curly line corresponds to a gluon.}
\label{fig:us}
\end{figure}

Because we are assuming that the binding energy, $E$, is much smaller than the temperature, we may integrate out energy scales of the order of $T$ as well. If we do so, we obtain an EFT which is temperature and velocity dependent. This EFT was called pNRQCD$_{\rm HTL}$ in \cite{Brambilla:2010vq},
where it was used in the case of vanishing velocity. We consider here the general case of non-vanishing velocity, following the analogous QED  calculation developed in \cite{Escobedo:2011ie}.
At the order we are considering, the pNRQCD$_{\rm HTL}$ Lagrangian is obtained from ${\cal L}_{\textrm{pNRQCD}}$ in Eq.~\eqref{pNRQCD} by replacing ${\cal L}_{\rm light} \rightarrow {\cal L}_{\rm vHTL}$, and $h_{s} \rightarrow h_{s}+\delta V_{s}$, where
${\cal L}_{\rm vHTL}$
is  the Hard Thermal Loop Lagrangian for a plasma moving with a velocity $v$ \cite{Weldon:1982aq}, and 
the potential  $\delta V_s$  encodes thermal contributions to the singlet potential, which depend on the velocity as well.
The expression of $\delta V_s$ can be obtained by a standard matching procedure, using dimensional regularization (DR) to regulate the IR divergences arising from the expansion $T\gg E$. In this case we have to consider the pNRQCD diagram in Fig.~\ref{fig:us}, where the dipole vertices and the octet propagator can be read off from \eqref{pNRQCD}, 	
\begin{equation}\label{deltaVs}
\delta V_s=-ig^2C_Fr_i\int\frac{\,d^Dk}{(2\pi)^D}\frac{i}{(E-h_o)-k_0+i\epsilon}(k_0^2D_{ij}(k)+k_ik_jD_{00}(k))r_j\,,
\end{equation}
with  $D_{\mu\nu}(k)$  the gluon propagator. Since $T\gg E$ and we use DR, the following expansion can be performed
\begin{equation}
\frac{i}{(E-h_o)-k_0+i\epsilon}=-i\left(\frac{1}{k_0-i\epsilon}+\frac{E-h_o}{(k_0-i\epsilon)^2}+\frac{(E-h_o)^2}{(k_0-i\epsilon)^3}+\cdots\right)\,,
\end{equation}
which corresponds to a temperature expansion, meaning that upon substituting this expression in Eq.~\eqref{deltaVs} we can expand the thermal contribution of the singlet potential as follows
\begin{equation}
\delta V_s=\delta V_{s,T^3}+\delta V_{s,T^2}+\delta V_{s,T}+\mathcal{O}(\alpha_s r^2 E^3)\,,
\end{equation}
where $\delta V_{s,T^n} \propto T^n$.
In the Coulomb gauge
\begin{equation}
\delta V_{s,T^3}=-g^2C_Fr^2\int\frac{\,d^Dk}{(2\pi)^D}k_0\left(\delta_{ij}-\frac{k_ik_j}{{\bm k^2}}\right)2\pi\delta(k_0^2-{\bm k}^2)f_B\!\!\left(\frac{|k_0-{\bm v}\cdot {\bm k}|}{\sqrt{1-v^2}}\right)\,,
\end{equation}
\begin{equation}
\delta V_{s,T^2}=-g^2C_Fr_i(E-h_0)r_j\int\frac{\,d^Dk}{(2\pi)^D}\left(\delta_{ij}-\frac{k_ik_j}{{\bm k^2}}\right)2\pi\delta(k_0^2-{\bm k}^2)f_B\!\!\left(\frac{|k_0-{\bm v}\cdot {\bm k}|}{\sqrt{1-v^2}}\right)\,,
\end{equation}
\begin{equation}
\delta V_{s,T}=-g^2C_Fr_i(E-h_0)^2r_j\int\frac{\,d^Dk}{(2\pi)^D}\frac{1}{k_0-i\epsilon}\left(\delta_{ij}-\frac{k_ik_j}{{\bm k^2}}\right)2\pi\delta(k_0^2-{\bm k}^2)f_B\!\!\left(\frac{|k_0-{\bm v}\cdot {\bm k}|}{\sqrt{1-v^2}}\right)\,,
\end{equation}
where $f_B(x)$ is the 
Bose-Einstein distribution function, see the Appendix \ref{appendix-gen}. Notice that $\delta V_{s,T^3}$ vanishes because the integrand is an odd function of $k_0$;  also  $\delta V_{s,T}$  vanishes because the integrand differs from an odd function by a $\delta (k_0)$, which forces the integral to be zero in DR. Hence, the only non-vanishing contribution is given by
\begin{equation}
\delta V_{s,T^2}=-\frac{g^2 C_F T^2}{12}r_i(E-h_o)r_j\left(P^s_{ij}+f(v)P^p_{ij}\right) \,,
\end{equation}
where we use the following notation, already introduced in \cite{Escobedo:2011ie},
\begin{equation}
f(v)=\frac{1}{v^3}\Big(v(2-v^2)-2(1-v^2)\tanh^{-1}(v)\Big) \,,
\end{equation}
\be
P_{ij}^s=\frac{1}{2}\left(\delta_{ij}+\frac{v_iv_j}{v^2}\right)\,, \qquad P_{ij}^p=\frac{1}{2}\left(\delta_{ij}-3\frac{v_iv_j}{v^2}\right)\,.
\ee
We can manipulate $r_i(E-h_0)r_j$ in the same way as it was done in the $v=0$ case in \cite{Brambilla:2010vq}, obtaining a more compact expression
\begin{equation}
\delta V_{s,T^2}=\frac{2\pi C_F \alpha_s T^2}{3m_Q}+\frac{\pi N_c C_F\alpha_s^2 T^2 r}{12}\left(1+f(v)+\frac{({\bm r}\cdot{\bm v})^2}{r^2v^2}(1-3f(v))\right)\,,
\end{equation}
where $N_c$ is the number of colors. The correction to the singlet potential in the pNRQCD$_{\rm HTL}$ Lagrangian is
\begin{equation}\label{eq:deltaVs}
\delta V_s=\frac{2\pi C_F \alpha_s T^2}{3m_Q}+\frac{\pi N_c C_F\alpha_s^2 T^2 r}{12}\left(1+f(v)+\frac{({\bm r}\cdot{\bm v})^2}{r^2v^2}(1-3f(v))\right)+
\mathcal{O}(\alpha_s r^2 E^3,\, \alpha_s^2 r^2 T^3)\,,
\end{equation}
where the $\mathcal{O}(\alpha_s^2 r^2 T^3)$ contributions above arise from $\alpha_s$ corrections to the diagram in Fig.~\ref{fig:us}. They have been calculated in \cite{Brambilla:2010vq} for the $v=0$ case.  
Differently from the hydrogen atom case \cite{Escobedo:2011ie}, the correction to the potential depends explicitly on the velocity. This could be expected from the fact that the Gromes relation (which is deduced by assuming Poincar\'{e} invariance) is violated at finite temperature \cite{Brambilla:2011mk}. However, as we shall detail in the next section,  for  the $s$-wave states the corresponding velocity dependence in the energy shifts  cancels out at first order in perturbation theory.

\subsection{Computation in pNRQCD$_{\rm HTL}$ and final results}\label{II-final}
With the obtained  pNRQCD$_{\rm HTL}$ Lagrangian we can evaluate  the thermal corrections to the binding energy and to the decay width of the various hydrogen-like states.  Since we are using perturbation theory,  we shall assume that the wave-functions are given by the unperturbed hydrogen atom solutions, which can be identified by the principal quantum number, $n$,  the angular momentum, $l$, and the magnetic quantum number, $m$. For a given heavy quarkonium state  the binding energy at the leading order in the perturbative expansion  is given by
\begin{equation}\label{perturbationth}
E_{nlm}=E_n^c+\langle n l m|\Re \delta V_s+\Re \Sigma_s|nlm\rangle\,,
\end{equation}
where $E_n^c$ is the binding energy taking into account only the Coulombic part of the potential and $\Sigma_s$ is the self-energy of the singlet component of the heavy quarkonium. Clearly, the Coulombic part of the potential does not contribute to  the decay width, which is nonzero only because of the thermal corrections, and at the leading order in the perturbative expansion it is given by 
\begin{equation}\label{widthn}
\Gamma_{nlm}=-2\langle nlm|\Im \delta V_s+\Im \Sigma_s|nlm\rangle \,.
\end{equation}
The singlet self-energy  can be determined computing the diagram in Fig.~\ref{fig:us} but this time in pNRQCD$_{\rm HTL}$. In order to properly take into account the moving thermal bath, the  boosted Bose-Einstein distribution function has to be used, and  since $T\gg E$, we expand
\begin{equation}\label{eq:expansionnB}
f_B\left(\frac{|k_0-{\bm v}\cdot{\bm k}|}{\sqrt{1-v^2}}\right)=\frac{T\sqrt{1-v^2}}{|k_0-{\bm v}\cdot{\bm k}|}+\cdots \,,
\end{equation}
hence the self-energy can be written as follows
\begin{equation}
\Sigma_s=-ig^2C_F r_i(E-h_o)^2\int\frac{\,d^Dk}{(2\pi)^D}\frac{T\sqrt{1-v^2}\delta(k_0^2-{\bm k}^2)}{|k_0-{\bm v}\cdot {\bm k}|}\frac{i}{(E-h_o)-k_0+i\epsilon}\left(\delta_{ij}-\frac{k_ik_j}{{\bm k}^2}\right)r_j +\mathcal{O}(\alpha_s r^2 E^3)\,.
\end{equation}
This integral is very similar to the one  evaluated in the QED case \cite{Escobedo:2011ie} and can be written as follows 
\begin{equation}
\Sigma_s=-ig^2C_Fr_i(E-h_o)^2\Re J_{ij}r_j\,,
\end{equation}
where $\Re J_{ij}$  is given in Eq. (48) of \cite{Escobedo:2011ie},
\be
\Re J_{ij}=\frac{T\sqrt{1-v^2}}{8\pi v}\left(P^s_{ij}\log\left(\frac{1+v}{1-v}\right)+P^p_{ij}\frac{\log\left(\frac{1+v}{1-v}\right)-2v}{v^2}\right) \,.
\label{ReJij}
\ee 
By manipulating $r_i(E-h_o)^2r_j$ we obtain the following result
\begin{eqnarray}
\Sigma_s &=&-\frac{i\alpha_s C_F T\sqrt{1-v^2}}{v}\left[\frac{p^2}{m_Q^2}\left\{\left(1+\frac{1}{v^2}\right)\log\left(\frac{1+v}{1-v}\right)-\frac{2}{v}+\left[\left(1-\frac{3}{v^2}\right)\log\left(\frac{1+v}{1-v}\right)+\frac{6}{v}\right]\frac{({\bm p}\cdot {\bm v})^2}{p^2v^2}\right\}\right.\nonumber \\
 &+&\left. \frac{N_c\alpha_s}{2m_Qr}\left\{\left(\frac{3}{2}-\frac{1}{2v^2}\right)\log\left(\frac{1+v}{1-v}\right)+\frac{1}{v}-\frac{1}{2}\left[\left(1-\frac{3}{v^2}\right)\log\left(\frac{1+v}{1-v}\right)+\frac{6}{v}\right]\frac{({\bm r}\cdot {\bm v})^2}{r^2v^2}\right\}\right.\nonumber \\
&+&\left. \frac{N_c^2\alpha_s^2}{16}\left\{\left(1+\frac{1}{v^2}\right)\log\left(\frac{1+v}{1-v}\right)-\frac{2}{v}+\left[\left(1-\frac{3}{v^2}\right)\log\left(\frac{1+v}{1-v}\right)+\frac{6}{v}\right]\frac{({\bm r}\cdot{\bm v})^2}{r^2v^2}\right\}\right]+\mathcal{O}(\alpha_s r^2E^3)\,,\label{eq:Sigmas}
\end{eqnarray}
and at the considered order it is pure imaginary. On the other hand, the correction to the singlet potential in Eq.~\eqref{eq:deltaVs} is real, with no imaginary part. Therefore, at the leading order
\be
\label{approx}
E_{nlm}=E_n^c+\langle n l m|\Re \delta V_s |nlm\rangle\,,  \qquad \Gamma_{nlm}=-2\langle nlm|\Im  \Sigma_s|nlm\rangle \,.
\ee
The correction to the binding energy in the  regime $1/r \gg T\gg E \gg m_D$ is given by
\begin{equation}\label{dEnlm}
\delta E_{nlm}=\frac{2\pi C_F T^2}{3} \left[\frac{\alpha_s}{m_Q}  + \frac{N_c \alpha_s^2}{2} \langle r \rangle_{nlm} +   \frac{N_c \alpha_s^2}{2} \langle r \rangle_{nlm} (1-3f(v))
\langle 2l00|l0\rangle\langle 2l0m|lm\rangle  \right] +\mathcal{O}(\alpha_s r^2E^3,\alpha_s^2 r^2 T^3 )\,,
\end{equation}
where  $\langle l''l'm''m'|lm\rangle$ are the Clebsch-Gordan coefficients and \be \langle r \rangle_{nlm} = \frac{a_0}{2}(3 n^2 -l(l+1))\,,\ee is the expectation value of the radial position operator in the hydrogen atom, with $a_0=1/m_Q C_F\alpha_s$ the Bohr radius.
In the expression of the binding energy we can distinguish three different contributions, corresponding to the three terms in the square bracket   of   Eq.~\eqref{dEnlm}. The first one is an overall energy shift, independent of the quantum state.  The second term  is a shift of the binding energy that removes the degeneracy in $l$ associated to the so-called ``accidental" symmetry of the hydrogen atom. The third term in the square bracket is an energy shift that depends not only on $n$ and $l$, but also on  $m$ and it is thereby related to the breaking of rotational invariance. There exists a privileged direction corresponding to $\bm v$, thus  the binding energy depends on the relative orientation between the angular momentum and  the velocity.    For $s$-wave states, having zero angular momentum, there is no such dependence  and this contribution to the binding energy vanishes, 
\begin{equation}
\delta E_n^{s-wave}=\frac{2\pi C_F\alpha_s T^2}{3m_Q}+\frac{\pi N_cC_F\alpha_s^2 T^2 a_0n^2}{2}+\mathcal{O}(\alpha_s r^2E^3,\alpha_s^2 r^2 T^3) \,,
\label{eq:caseIenergy}
\end{equation}
and as already anticipated it does not depend  on the velocity of the plasma. The latter result is surprising, because one would have naively expected that  a $v$ dependence should arise  because  for the moving bound state the effective temperature depends on $v$, see the Appendix \ref{appendix-gen}. However, our calculation shows that this is not the case, at least at the first order in perturbation theory.

Regarding the width,  from the expression of the self energy in Eq.~\eqref{eq:Sigmas}, we obtain
\begin{eqnarray}\label{eq:gammanlm}
\Gamma_{nl m}&=&\frac{\alpha_s C_FT\sqrt{1-v^2}}{3v}\left[4\left(-\frac{2E_n^c}{m_Q}+\frac{\alpha_s N_c}{m_Qa_0 n^2}+\frac{\alpha_s^2 N_c^2}{8}\right)\log\left(\frac{1+v}{1-v}\right)+\left(-\frac{4E_n^c}{m_Q}-\frac{\alpha_s N_c}{m_Qa_0n^2}+\frac{\alpha_s^2 N_c^2}{4}\right)h_{lm}(v) \right]\nonumber \\
&+&\mathcal{O}(\alpha_s r^2E^3,\alpha_s^2 r^2 T^3)\,,
\end{eqnarray}
where
\begin{equation}
h_{lm}(v)=\left[\left(1-\frac{3}{v^2}\right)\log\left(\frac{1+v}{1-v}\right)+\frac{6}{v}\right] \langle 2l00|l0\rangle\langle 2l0m|lm\rangle\,,
\end{equation}
is a negative and  decreasing function of $v$. It can be easily  shown, using the  expression above,  that for any state the width  is a decreasing function of the velocity,  vanishing for $v \rightarrow 1$, meaning that  an ultra-relativistic velocity has the effect of stabilizing the system. This behavior   is due to the $\sqrt{1-v^2}$ prefactor in Eq.~\eqref{eq:gammanlm},  which can be traced back to the expansion of the boosted Bose-Einstein distribution function in Eq.~\eqref{eq:expansionnB} and is therefore due to the ``Doppler shift" of the temperature, see the Appendix \ref{appendix-gen}.  As we shall see in the next section, an analogous behavior is obtained for  the hierarchy of energy scales of  Case II in the ultra-relativistic limit, although the microscopic description appears to be different.

 In the expression of the width  we can further distinguish two contributions, corresponding to the two terms in the square bracket in Eq.~\eqref{eq:gammanlm}. Both are velocity dependent, but the first  one does only  depend on $n$, meaning that it originates from terms that do not break the rotational and the accidental symmetries. The second term depends on all the quantum numbers and vanishes for $s$-wave states. Thus, for  $s$-wave states the width simplifies to
\begin{equation}
\Gamma_n^{s-wave}=\frac{4\alpha_s C_FT\sqrt{1-v^2}}{3v}\left(-\frac{2E_n^c}{m_Q}+\frac{\alpha_sN_c}{m_Qa_0n^2}+\frac{\alpha_s^2 N_c^2}{8}\right)
\log\left(\frac{1+v}{1-v}\right)+\mathcal{O}(\alpha_s r^2E^3,\alpha_s^2 r^2 T^3)\,,
\label{eq:width-swave}\end{equation}
which, as observed above for the general case, is a decreasing function of the velocity, vanishing for $v \rightarrow 1$.

\section{Case II}\label{sec:CaseII}
The dissociation of heavy quarkonium is expected to occur for
\be
T \gg 1/r \,, m_D \gg E\,,
\ee
 as it happens in a thermal bath at rest. In the color-screening model introduced in \cite{Matsui:1986dk} the dissociation takes place because the number of bound states supported by a Yukawa potential decreases with the range of the potential, which is proportional to the screening length ($1/m_D$). 
The effect of screening becomes important when the screening length is of the order of the size of the system ($1/m_D\sim r$). However in the actual real-time potential computed in \cite{Laine:2006ns} 
(and confirmed by the EFT computations \cite{Escobedo:2008sy,Brambilla:2008cx}) the dissociation takes place because the potential develops an imaginary part (Landau damping), and bound states turn into wide resonances as the temperature increases. This effect becomes important at a parametrically different scale, $1/(Tm_D^2)^{1/3}\sim r$ (up to logarithms) \cite{Escobedo:2008sy}.  
It is then particularly interesting to address the question whether the mechanism of dissociation (screening versus Landau damping) remains the same when the bound state moves with respect to the thermal bath.

An EFT study of this situation in the QED case for muonic hydrogen submerged in a bath of massless electrons
was already performed in \cite{Escobedo:2011ie}(see \cite{Escobedo:2010tu} for the $v=0$ case). For heavy quarkonium the results are analogous and can be obtained by changing the value of the Debye mass from the QED to the QCD value and by correcting for trivial color factors, as it was already pointed out in \cite{Escobedo:2011ie}. We briefly review the basic steps of the derivation 
below.
\begin{itemize}
\item Since $m_Q\gg T$, the starting point of the calculation can be the NRQCD Lagrangian at zero temperature \cite{Caswell:1985ui}. By integrating out the temperature scale we arrive at the NRQCD$_{\rm HTL}$, an EFT whose Lagrangian is the sum of NRQCD for the heavy quark sector (with thermal, velocity independent, corrections to the heavy quark mass) \cite{Escobedo:2008sy}, and the HTL Lagrangian for gluons and light quarks, which now depends on the relative velocity $v$ between the thermal bath and the bound state \cite{Weldon:1982aq}.
\item We can also integrate out the scales $1/r$ and $m_D$, which leads to pNRQCD$_{\rm HTL}$.  
Since gluons and light quarks in the HTL Lagrangian develop a mass gap of the order of $m_D$, this effective theory does not contain them as explicit degrees of freedom, and hence it reduces to a singlet heavy quark-antiquark field interacting through a potential $V_s$ (which depends on $v$ as well). 
The main conceptual difference with respect to the case discussed in the previous section is that now, in general, the 
thermal contributions cannot be considered as a perturbation in the potential.  
\end{itemize}
The potential $V_s$ coincides with the one that was computed numerically in Sec.V of \cite{Escobedo:2011ie}. In that paper qualitative arguments were put forward on how the dissociation mechanism is modified when the velocity of the bound state with respect to the plasma increases. 
We will quantify those arguments here, by focusing on the effects of this potential on the physics of the 1$s$ state. We shall discuss two different cases.
\begin{enumerate}
\item We consider the particular case $1/r\gg m_D$. The thermal contributions can still be considered as a  perturbation to the Coulomb potential. This allows us to compute the leading thermal corrections to the decay width (almost) analytically and to derive some explicit expressions for the velocity dependence. In particular we can parametrically estimate how the dissociation temperature depends on the velocity if $1-v \gg m_Da_0$. 
Then, as in the $v=0$ case, the dissociation mechanism is dominated by Landau damping effects. 
This has to be contrasted to Case I in Sec.~\ref{sec:CaseI} where the decay width is entirely due to gluo-dissociation. 
\item We consider the general case in which the relative size between $1/r$ and $m_D$ is left arbitrary ($1/r\sim m_D$) and  compute the spectral function. Although the concept of dissociation temperature is useful for qualitative estimates, there is no universal definition for it, and hence it is of limited usefulness for a quantitative comparison of our results with other approaches. On the contrary, the spectral function is a well defined quantity so that our results  can be straightforwardly compared with those obtained by different  methods, in particular by lattice computations. Furthermore, it is related to a physical observable, the thermal  dilepton production rate \cite{McLerran:1984ay,Weldon:1990iw}.
In current HIC experiments, however, the heavy quarkonium states are not expected to be thermalized, but rather to act as hard probes of the medium, and hence the connection of the spectral function to the dilepton spectrum in this case is not straightforward.
In the spectral function a bound state with zero decay width appears as a delta function whereas scattering states produce a smooth curve. The spectral function allows us to observe all the intermediate situations which occur when changing the  thermal bath temperature and velocity. 
\end{enumerate}

For the expressions of the coupling constant and the Debye mass, we will use the following parameterization (we set  $N_c=3, N_f=3$),
\be
\alpha_s=\alpha_s(1/a_0) 
\qquad m_D =\frac{2\,\pi\, T}{\sqrt{3\log(2\pi\, T/\Lambda_{\overline{\rm MS} })}}=\sqrt{4\pi\alpha_s(2\pi T)T^2(N_c+N_f/2)}\,,
\ee
where  the $\overline{\rm MS}$  renormalization scheme has been adopted with $\Lambda_{\overline{\rm MS} }=250$ MeV; we also fix $m_Q=4.881$ GeV and the Bohr radius of $\Upsilon(1s)$ is given by $a_0 \simeq 0.74$ GeV$^{-1}$, both values are taken from \cite{Pineda:1998id}. This choice is motivated from the fact that   computing higher order corrections to the potential would introduce a dependence on the renormalization scale of the type $\log^n(r\mu)$; on the other hand, computing higher order corrections to the Debye mass would introduce terms proportional to $\log^n(2\pi T/\mu)$.

\subsection{The particular case $T\gg 1/r\gg m_D\gg E$}\label{sec:particular}
In this case the potential can be considered as the Coulomb potential plus a perturbation, and hence the following formula provides a good approximation to the decay width of a $s$-wave state
\begin{equation}
\Gamma_{n}^{s-wave}=-2\langle n00|\Im V_s(r)|n00\rangle=-\int\,d^3{\bm r}|\psi_{n}({\bm r})|^2\int\frac{\,d^3{\bm k}}{(2\pi)^3}(e^{i{\bm k}\cdot {\bm r}}-1)\Delta_S({\bm k},{\bm v})\,,
\label{eq:gamma1s}
\end{equation}
where $\psi_{n}({\bm r})$ is the wave-function for a $s$-wave state in the Coulomb potential and $\Delta_S$ is the symmetric part  of the $00$ component of the gluon field  propagator in the Coulomb gauge, which has been computed in \cite{Escobedo:2011ie} for QED. Its generalization to QCD can be straightforwardly obtained by introducing a color factor $C_F$ and substituting the value of $m_D$ by the corresponding QCD one,
\begin{equation}
\Delta_S({\bm k},{\bm v})=\frac{8\pi^2\alpha_s C_F Tm_D^2f(v,\theta)}{k(k^2+m_D^2g(z,v))(k^2+m_D^2g^*(z,v))}\,,
\end{equation}
and depends on $v$, $k=\vert {\bm k}\vert$ and on $\theta$, the angle between the vectors ${\bm k}$ and ${\bm v}$. In the above equation
\begin{equation}
f(v,\theta)=\frac{(1-v^2)^{3/2}(2+v^2\sin^2\theta)}{2(1-v^2\sin^2\theta)^{5/2}}\,,
\end{equation}
and we have made the dependence on the Debye mass explicit by defining $g(z,v)=\Pi_R(z,v)/m_D^2$, with \begin{equation}
z=\frac{v\cos\theta}{\sqrt{1-v^2\sin^2\theta}}\,, 
\end{equation} and $\Pi_R(z,v)$ is the retarded self-energy of the $00$ 
component of the gluon field in the Coulomb gauge with $k_0=0$, which was first computed in~\cite{Chu:1988wh}\footnote{In \cite{Chu:1988wh} there is a misprint in the first line of Eq.~(8), in which the global sign must be the 
opposite.}.  In the reference frame where the bound state is at 
rest
\begin{equation}
\Pi_R(z,v)=a(z)+\frac{b(z)}{1-v^2},
\end{equation}
with 
\begin{equation}
a(z)=\frac{m_D^2}{2}\left[z^2-(z^2-1)\frac{z}{2}\ln\left(\frac{z+1+i\epsilon}{z-1+i\epsilon}\right)\right],
\end{equation}
and
\begin{equation}
b(z)=(z^2-1)\left[a(z)-m_D^2(1-z^2)\left(1-\frac{z}{2}\ln\left(\frac{z+1+i\epsilon}{z-1+i\epsilon}\right)\right)\right]\,.
\end{equation}

In principle we can obtain the decay width for any $s$-wave state, but for illustrative purposes we shall focus on the ground state ($n=1$). It is convenient to start the computation in Eq.~\eqref{eq:gamma1s} by first performing the integration over $r$,
\begin{equation}
\Gamma_{1}^{s-wave}=2\alpha_s C_F Tm_D^2\int_{-1}^1d\cos\theta f(v,\theta)\int_0^\infty\frac{\,dkk}{(k^2+m_D^2g(z,v))(k^2+m_D^2g^*(z,v))}\left(1-\frac{1}{(1+\frac{k^2a_0^2}{4})^2}\right)\,,
\label{eq:gammatot}
\end{equation}
where we have switched to cylindrical coordinates in momentum space and performed as well the trivial integration over the azimuthal angle. 

The above expression can be numerically integrated, but we first obtain an approximate expression valid at
moderate velocities. In this case, for any angle, $g(z,v)$ is of order $1$ and there are  only two scales in the previous 
integral, $1/a_0$ and $m_D$. Moreover they fulfill the relation $1/a_0\gg m_D$, so that the technique of threshold expansion \cite{Beneke:1997zp} can be used to work out the integral, thus obtaining
\begin{eqnarray}
\Gamma_{1}^{s-wave}&=&\alpha_s C_FTm_D^2a_0^2 
 \int_{-1}^1d\cos\theta \, f(v,\theta)\left(\log\left(\frac{2}{m_Da_0}\right)-\frac{1}{4}-\frac{g(z,v)\log(g(z,v))-g^*(z,v)
\log(g^*(z,v))}{2(g(z,v)-g^*(z,v))}\right. \nonumber  \\
 &+& \left. {\cal O}\left((m_Da_0)^2\right)\right) 
\,.
\label{eq:gammathr}
\end{eqnarray}
This equation can be further simplified by taking into account that the $\log\left(\frac{2}{m_Da_0}\right)$ is logarithmically bigger than the rest of 
the terms in the parenthesis. With this approximation we arrive at the following result,
\begin{equation}
\Gamma_{1}^{s-wave}\sim\frac{2\alpha_sC_FTm_D^2a_0^2}{\sqrt{1-v^2}}\log\left(\frac{2}{m_Da_0}\right)\,,
\end{equation}
or, equivalently,
\begin{equation}
\frac{\Gamma_{1}^{s-wave}(v)}{\Gamma_{1}^{s-wave}(v=0)}\sim\frac{1}{\sqrt{1-v^2}}\,,
\label{eq:gammav}
\end{equation}
which holds up to ${\cal O}(1/\log (m_Da_0))$ accuracy and is independent of the heavy quark mass and of the temperature. Then, in the regime $T\gg 1/r\gg m_D$, the decay width increases with the velocity, as far as it remains moderate
($v\nsim 1$).  Note that this behavior is opposite to the one observed in the regime $1/r \gg T \gg E \gg m_D$ in Sec. \ref{II-final}, see  Eq.~\eqref{eq:width-swave}. If we take into account that $m_D^2$ is proportional to $T^2$, 
the decay width at temperature $T$ and  velocity $v$ is the same as the one that we would observe at $v=0$ but with 
\begin{equation}
T \rightarrow T_v=\frac{T}{(1-v^2)^{1/6}}\,,
\end{equation}
provided that $T_v\sim T\gg 1/r\gg m_D$.

\begin{figure}[b!]
\subfigure{
 \includegraphics[width=8cm]{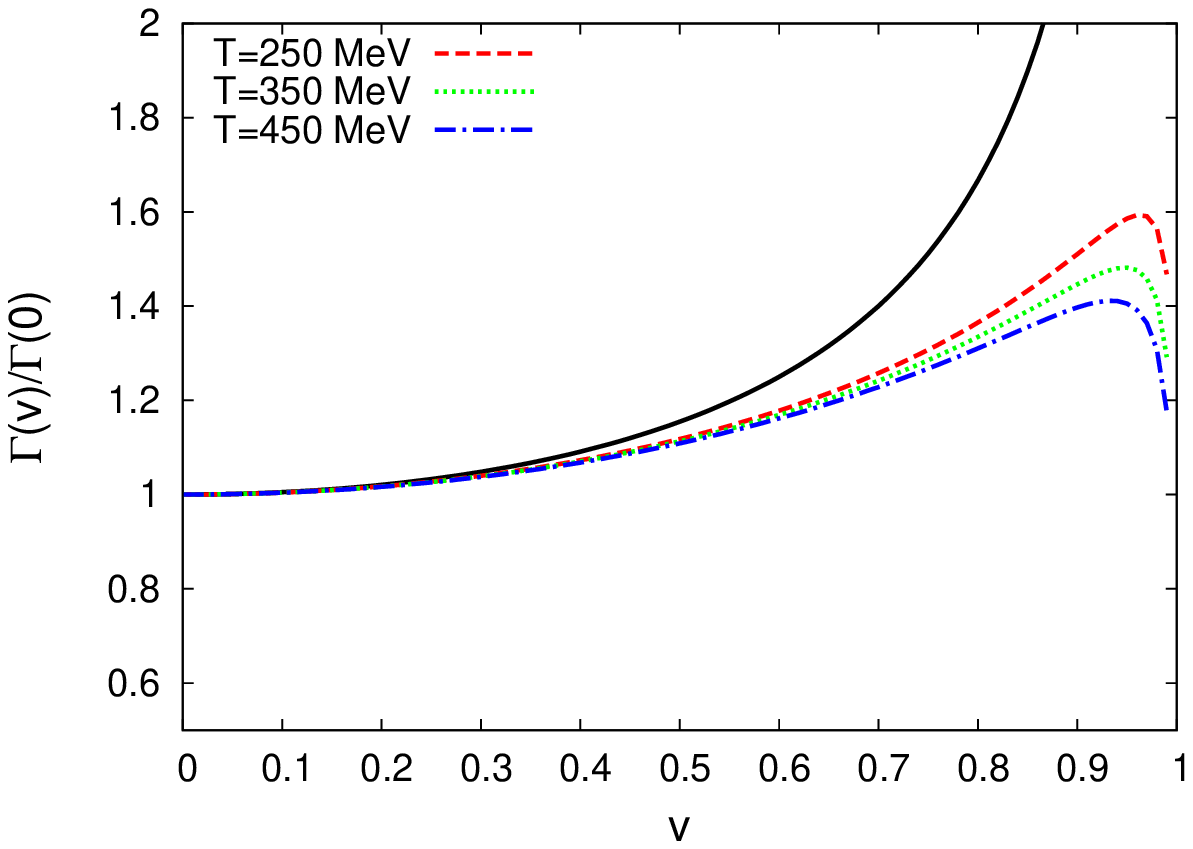}}
\subfigure{
\includegraphics[width=8cm]{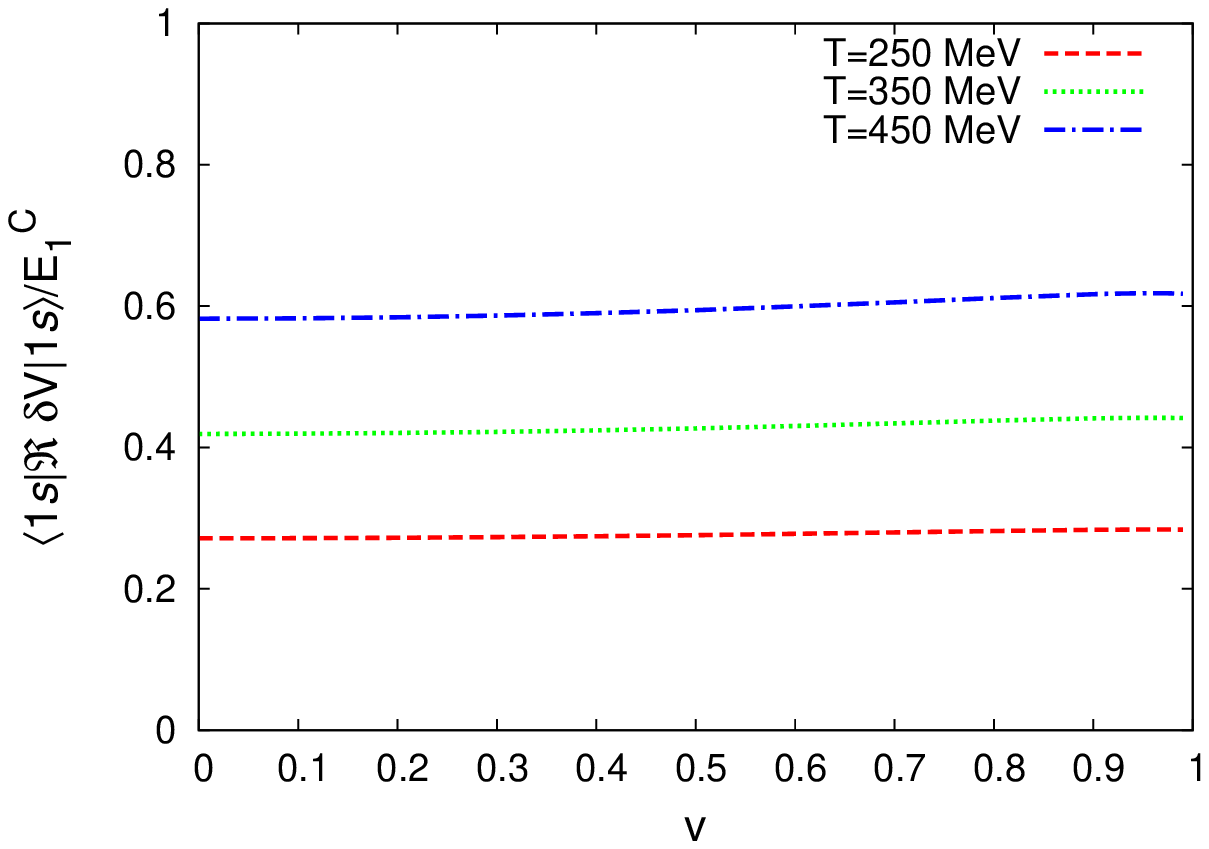}}
  \caption{(Color online) Left panel: Width $\Gamma(v)/\Gamma(0)$ from Eq.~\eqref{eq:gammav} (solid line) and  $\Gamma(v)/\Gamma(0)$ of the $\Upsilon(1s)$  from Eq.~\eqref{eq:gammatot} at a few values of temperature, as a function of the velocity $v$. Right panel: Expectation value of the real part of $\delta V/E_1^c$ as a function of the velocity, $\delta V$ being the thermal contribution to the singlet potential and $E_1^c$ the Coulomb binding energy for the 1$s$ state. }
 \label{fig:gamma}
\end{figure}

Results beyond the logarithmic accuracy of \eqref{eq:gammav} can be obtained by evaluating \eqref{eq:gammathr}, or even better \eqref{eq:gammatot}. The numerical values of $\Gamma(v)/\Gamma(0)$ for the $\Upsilon(1s)$ state are   reported  in the left panel of Fig.~\ref{fig:gamma}, for three different temperatures, together with the approximate expression \eqref{eq:gammav}, solid black line,  for $0 \le v \lesssim 1$.  The approximate expression  correctly reproduces the  numerical values for  $v \lesssim 0.5$, but for larger values of $v$   the ratio of the width decreases and becomes temperature dependent, departing from Eq.\eqref{eq:gammav}. This is due to the fact that for $v \rightarrow 1$ further scales must be 
considered ($T_+\gg T\gg T_-$) and  \eqref{eq:gammathr} does not hold.  This expression 
relies on the fact that $1/a_0^2 \gg m_D^2 |g(z,v)|$, which not always holds for $v\to 1$, even if $1/a_0^2 \gg m_D^2$ does.

\begin{figure}[ht!]
\subfigure{ \includegraphics[width=8cm]{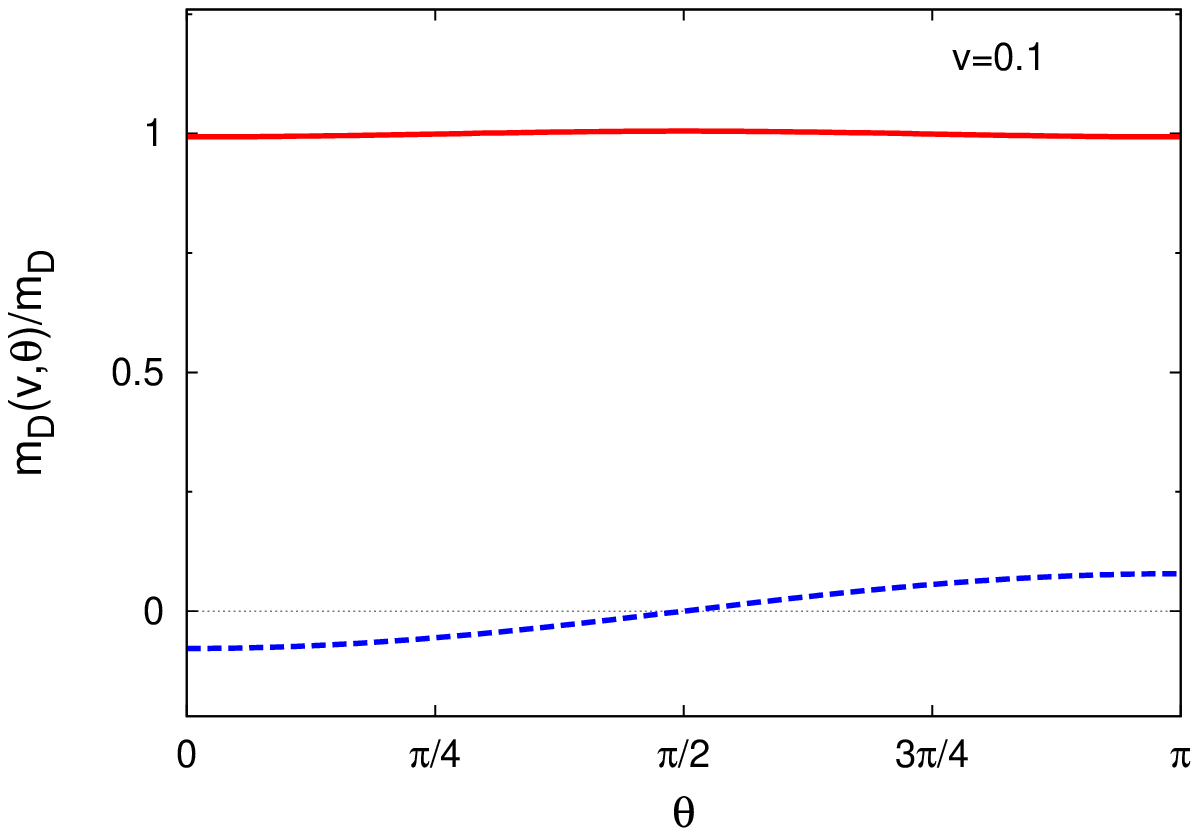}}
\subfigure{ \includegraphics[width=8cm]{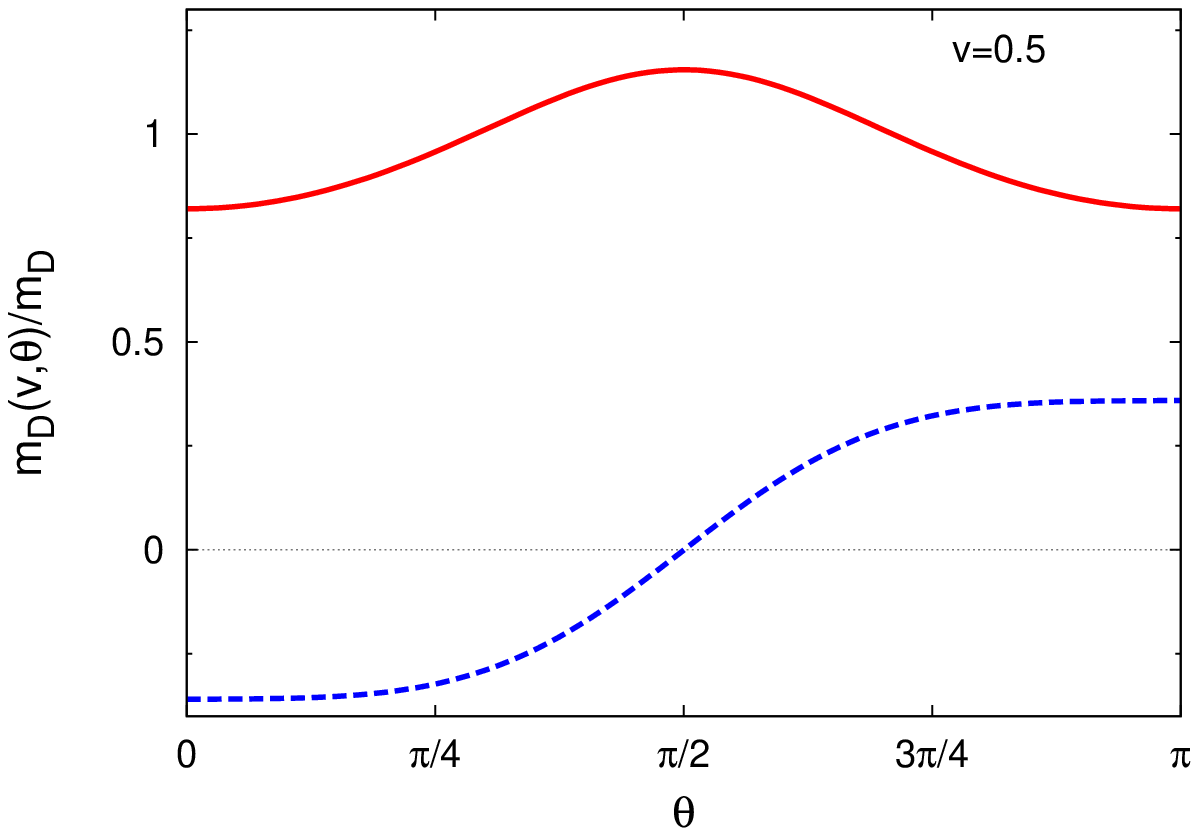}}
\subfigure{ \includegraphics[width=8cm]{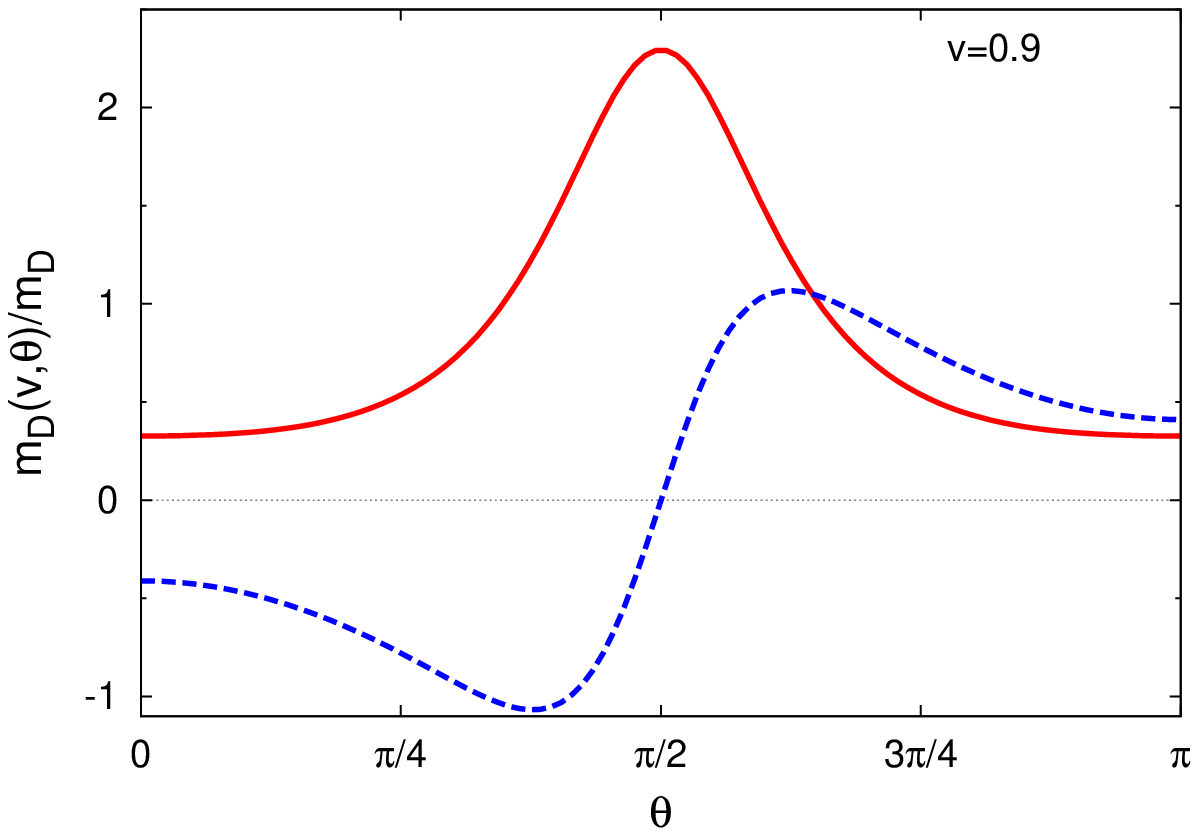}}
\subfigure{ \includegraphics[width=8cm]{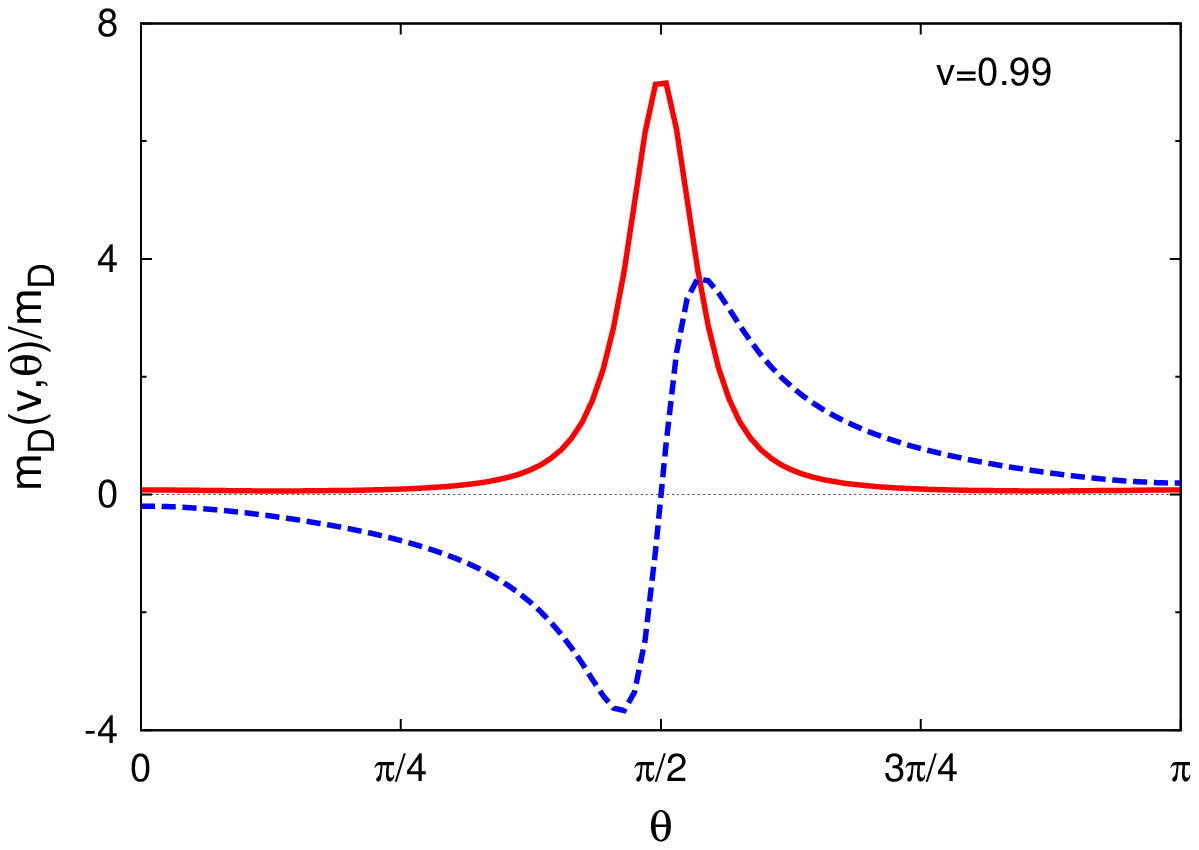} }
  \caption{(Color online) Real (solid red) and imaginary (dashed blue) parts of the velocity dependent Debye mass as a  function of the angle between the vectors ${\bm k}$ and ${\bm v}$, for $v=0.1, 0.5, 0.9, 0.99$. }
 \label{fig:debye-mass}
\end{figure}

In order to ascertain the reliability of the expression in \eqref{eq:gammatot}  and the origin of the difference between \eqref{eq:gammatot} and  \eqref{eq:gammav}, let us scrutinize the 
velocity and angular dependence of $m_D^2 g(z,v)$. 
We can interpret the square root of  $m_D^2 g(z,v)$ with positive real part \cite{Chu:1988wh, Escobedo:2011ie}  as the velocity  dependent Debye mass $m_D(v,\theta)$  ($m_D(v,\theta)$ should not be mistaken for the parameter $m_D$ that we used  before, they coincide at $v=0$ only and have the same size for moderate velocities $v\nsim 1$ only).  In Fig.~\ref{fig:debye-mass} we present the plots of  $\Re [m_D(v,\theta)]/ m_D$ and of $\Im [m_D(v,\theta)]/ m_D$ as a function of $\theta$ for $v=0.1, 0.5, 0.9, 0.99$. 
 The real part is peaked at $\theta = \pi/2$, corresponding to a vanishing   value of the imaginary part, which is instead peaked at a  value of $\theta$  that with increasing $v$ approaches $\pi/2$. 
 For $\theta \nsim \pi/2$,  $m_D(v,\theta)$  is small and the imaginary part is 
of  ${\cal O}(m_D)$ for any value of $v$, meaning that the  bound state can be approximated with a Coulombic wave function, and therefore in this region \eqref{eq:gamma1s} represents a good approximation. Moreover, for $v \lesssim 0.5$ both  the real and the imaginary part of  $m_D^2 g(z,v)$ are of ${\cal O}(m_D)$, irrespective of the value of $\theta$, and therefore the approximate expression  \eqref{eq:gammav} is reliable. This approximation is still qualitatively good up to $v \simeq 0.9$, although the increased value of $m_D(v,\theta)$ for $\theta \sim \pi/2$, suggests that the quantitative  agreement might be lost, as indeed can be observed in the left panel of Fig. \ref{fig:gamma}.

An angular region that may jeopardize the perturbative expansion about the Coulombic wave function 
is only present for $v > 0.9$, around $\theta \sim \pi/2$.  Indeed for $\theta=\pi/2$ the real part of the Debye mass has a peak,  and for $\theta \sim \pi/2$ the imaginary part is large. However, this angular region 
is small. In order to quantify this   region we consider  $\Im[m^2_D(v,\theta )] = 2 \Re[m_D(v,\theta) ] \Im[m_D(v,\theta) ]$, which has two maxima for 
\be
\theta_{\pm} = \arccos\left[ \pm \frac{1}{2 v} \sqrt{11 + v^2 -\sqrt{3}\sqrt{35 + 10 v^2 + 3 v^4}}\right]\,,
\ee
and therefore for $v > 0.9$ the angular region around $\theta=\pi/2$ where the real and the imaginary parts are large is given by 
\be
\Delta\theta = |\theta_+ - \theta_-| \simeq \sqrt{2(1-v)} \label{deltatheta}\,,
\ee
which clearly  shrinks to zero for $v \rightarrow 1$. In order to clarify that the contribution of this angular region is small,
 we plot in the right panel of Fig.~\ref{fig:gamma} the thermal correction to the Coulomb binding energy $\langle 100|\Re \delta V(r)|100\rangle$, $\delta V$ being the thermal contribution to the singlet potential, normalized to the Coulomb binding energy. This quantity should be small for \eqref{eq:gammatot} to be reliable, as it turns out to be the case.  
  
Note, however, that  the angular  region $\Delta\theta$ gives for $v\to 1$ the largest contribution to \eqref{eq:gammatot}, and in this case $g(z,v)\sim b(z)/(1-v^2)$.  Since $m_D^2 |g(z,v)|\gg 1/a_0^2$,  the approximate expression in \eqref{eq:gammathr} does not hold anymore, in agreement with the results in the left panel of Fig.~\ref{fig:gamma}. We find in this case, using the same techniques of integration by regions, that for $v \rightarrow 1$ the decay width goes to zero like $\alpha_s T \sqrt{1-v^2}$ whereas the energy shift goes to a constant, consistent with the results reported in the right panel of Fig.~\ref{fig:gamma}. 

The velocity  dependence of the width in the ultra-relativistic limit is similar to the one discussed in the Case I, in Sec.\ref{II-final}, although the microscopic mechanism is different, Landau damping in the present case, and gluo-dissociation in the former. The reason for the decrease in the decay width is probably related to the fact that a moving bound state   feels a plasma with a non-isotropic  \textit{effective temperature}
\be\label{effective-temperature}
T_{\rm eff}(\theta,v)=\frac{T\sqrt{1-v^2}}{1-v\cos{\theta}}\,,
\ee
 see Appendix \ref{appendix-gen} for more details. Actually, the effective temperature is higher than $T$ in the forward direction, and lower than $T$ in the backward direction, thus it is not obvious that the width of the moving bound state should increase --- or be modified at all ---  when the bound state moves with respect to the thermal medium. However, in the ultra-relativistic case the effective temperature is almost everywhere less than $T$, see Fig.~\ref{fig:effectiveT}, and it is higher than $T$ only in a narrow region $\theta \sim 0$. According to the previous discussion this angular region  does not give the leading contribution to the width, which is instead dominated by the $\theta \sim \pi/2$ region, where the heat bath is effectively cold. Thus, a velocity close to $1$ tends to stabilize the system.

The presence of an imaginary Debye mass for any $v >0$ can be related to the collisionless transfer of energy between the heavy quarks and the  gauge fields. An accurate description of this phenomenon would require the discussion of the propagating modes, but  an imaginary part of  the Debye mass does in any case signal an instability.
This phenomenon is akin to  the plasma instabilities generated by a charged current in a plasma. A similar result was indeed obtained in \cite{Mannarelli:2007gi, Mannarelli:2007hj} where  the destabilizing effect of a  single heavy quark propagating in a thermalized QGP was studied.

\subsection{ The general case $T \gg 1/r$ , $m_D \gg E$: the spectral function} \label{sec:CaseII_spectral}

The $s$-wave spectral function was computed in \cite{Laine:2007gj,Burnier:2007qm} for the case at rest ($v=0$).
The procedure developed in Ref.~\cite{Laine:2007gj} can be easily  generalized to a moving bound state. We shall use  the expression of the potential determined in~\cite{Escobedo:2011ie}, which takes into account the 
relative velocity,  $v$, between the bound state and  the expanding plasma, and consider that the system  has  cylindrical symmetry, with its symmetry 
axis in the direction of $\bm v$.

The formalism introduced in \cite{Laine:2007gj} can be generalized to  cylindrical coordinates, resulting in the following expression of the spectral function 
\begin{equation}\label{eq:specfun}
\rho(\omega)=\lim \limits_{\substack{r\rightarrow 0\\z\rightarrow 0}}  \int_{0}^{+\infty} dt\, \left\{ \cos(\omega t)\, \Re[\psi(t,r,z)]-\sin(\omega t)\, \Im[\psi(t,r,z)] \right\} \,,
\end{equation}
where $\psi(t,r,z)=u(t,r,z)/r$, and $u(t,r,z)$ is the solution of the Schr\"odinger equation
\begin{equation}\label{eq:Scheq}
\left(i\partial_t+\frac{1}{m_Q}\frac{\partial^2}{\partial^2z}+\frac{1}{m_Q}\frac{\partial^2}{\partial^2r}+\frac{1}{m_Qr^2}-\frac{1}{m_Qr}\frac{\partial}{\partial r}-V(r,z)\right)u(t,r,z)=0 \,,
\end{equation}
with the initial condition $u(0,r,z)=-6 N_c r \delta^2(r) \delta(z)$.

In order to numerically handle the two-dimensional Schr\"odinger  equation, we use an {\it operator-splitting} method, namely we  split the differential equation  in two differential equations each containing   derivatives with respect to only one variable ($z$ or $r$).  Therefore, the whole Hamiltonian is divided in two pieces, $H=H_1+H_2$, where
\bea
H_1&=& -\left(\frac{1}{m_Q}\frac{\partial^2}{\partial^2z}-V(r,z)\right) \,,\\
H_2&=& -\left(\frac{1}{m_Q}\frac{\partial^2}{\partial^2r}+\frac{1}{m_Qr^2}-\frac{1}{m_Qr}\frac{\partial}{\partial r}\right)\,,
\eea
and we then solve the corresponding Schr\"odinger equation recursively in a discrete space-time,  see the Appendix \ref{appendix-osm} for more details 
and for a check of the numerical code.

\begin{center}
\begin{figure}[h!]

\subfigure{
 \includegraphics[width=8cm]{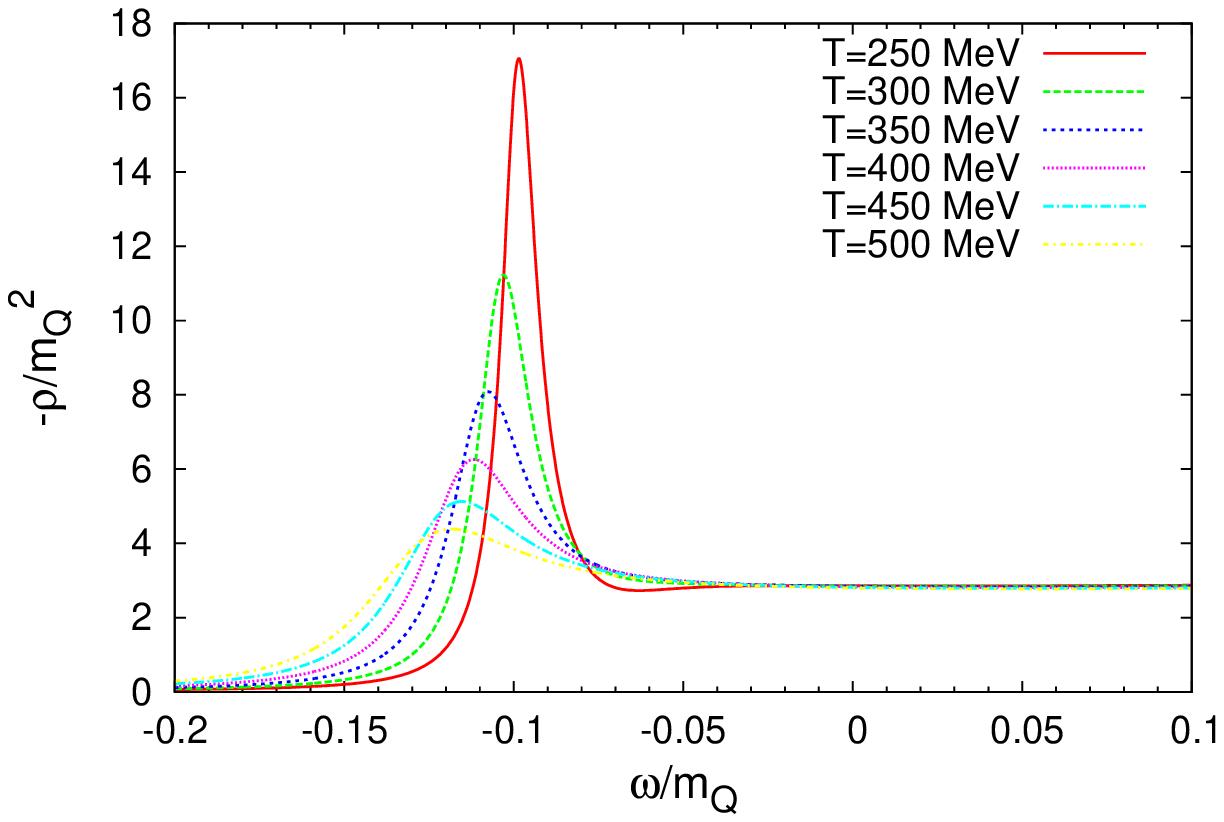}}
\subfigure{
\includegraphics[width=8cm]{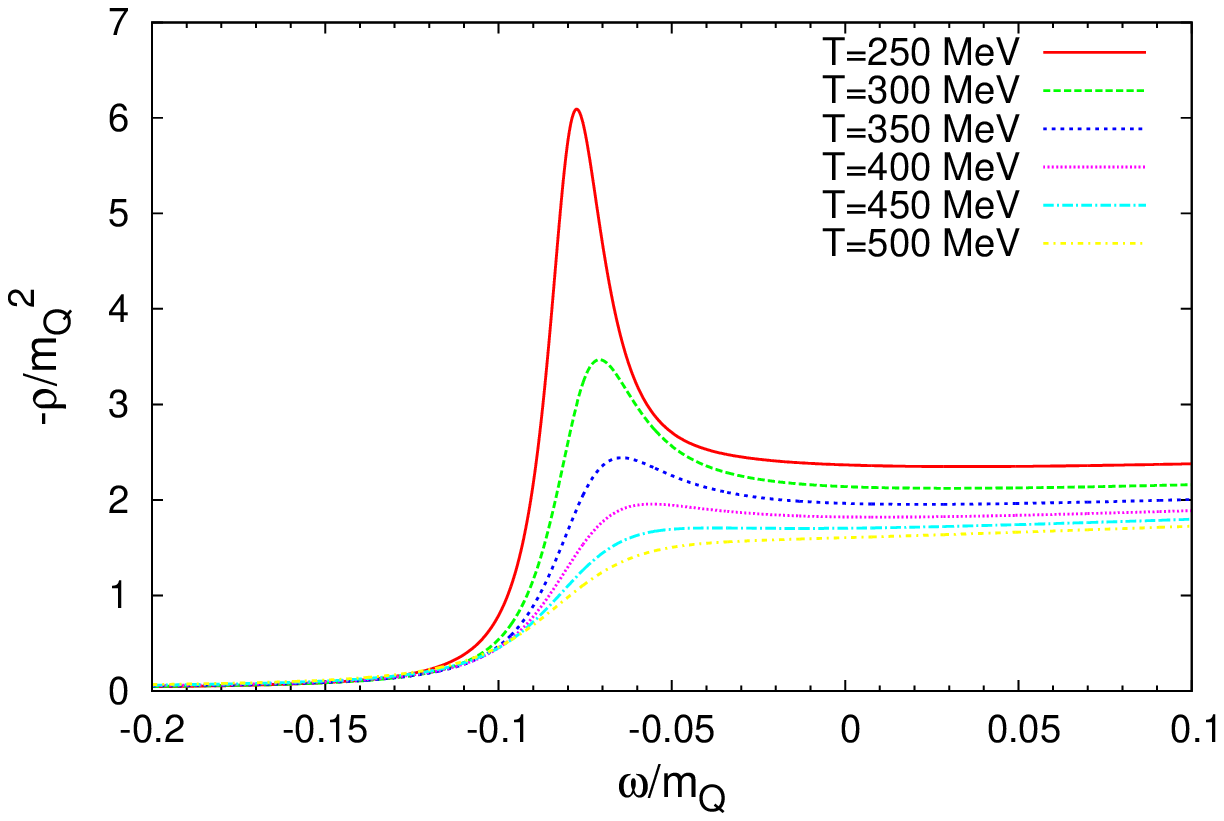}}
 \caption{(Color online) Spectral function of the $\Upsilon(1s)$  state, divided by $-m_Q^2$, at $v=0$ 
 for our choice of scales for $\alpha_s$ (left panel) and for the choice of Refs.~\cite{Kajantie:1997tt,Laine:2007gj} (right panel). }
 \label{fig:SFv0}
\end{figure}
\end{center}

 In Fig.~\ref{fig:SFv0} we report the 
spectral functions at vanishing  velocity for certain values of the temperature as a function of $\omega/m_Q$, where $\omega \ll m_Q$ is the non-relativistic energy ($\omega=0$ corresponds to a relativistic energy of $2m_Q$). For the sake of comparison, in the right panel of Fig.~\ref{fig:SFv0} we also report the spectral functions obtained in \cite{Laine:2007gj} with a different choice of $\alpha_s$ and $m_D$, see Refs.~\cite{Kajantie:1997tt,Burnier:2007qm,Laine:2007gj} for more details. 
At $T = 250$ MeV the spectral function is given by the superposition of a peak, corresponding to the $\Upsilon(1s)$ bound state  and of a continuum. 
The bound state has a thermal width which is determined by the imaginary part of the potential and is dominated by the Landau damping. The width of the bound state increases with increasing temperature, and correspondingly, the contribution of the continuum increases. At $T \simeq 500$ MeV no peak of the spectral function is visible, meaning that   the bound state has dissolved into the continuum.  

With increasing temperature the position of the peak slightly drifts away to the left, meaning that the $\Upsilon(1s)$ mass decreases. The binding energy decreases as well, since the absolute value of the real part of the potential for $r,z\to \infty$ increases. In fact, if one subtracts such an asymptotic value from $\omega$, in both figures the peak drifts to the right as the temperature increases. However, the bound state does not disappear because the binding energy vanishes, but because for $T \ge 400$ MeV Landau damping prevents the formation of a bound state.

\begin{figure}[h!]
\subfigure{
 \includegraphics[width=8cm]{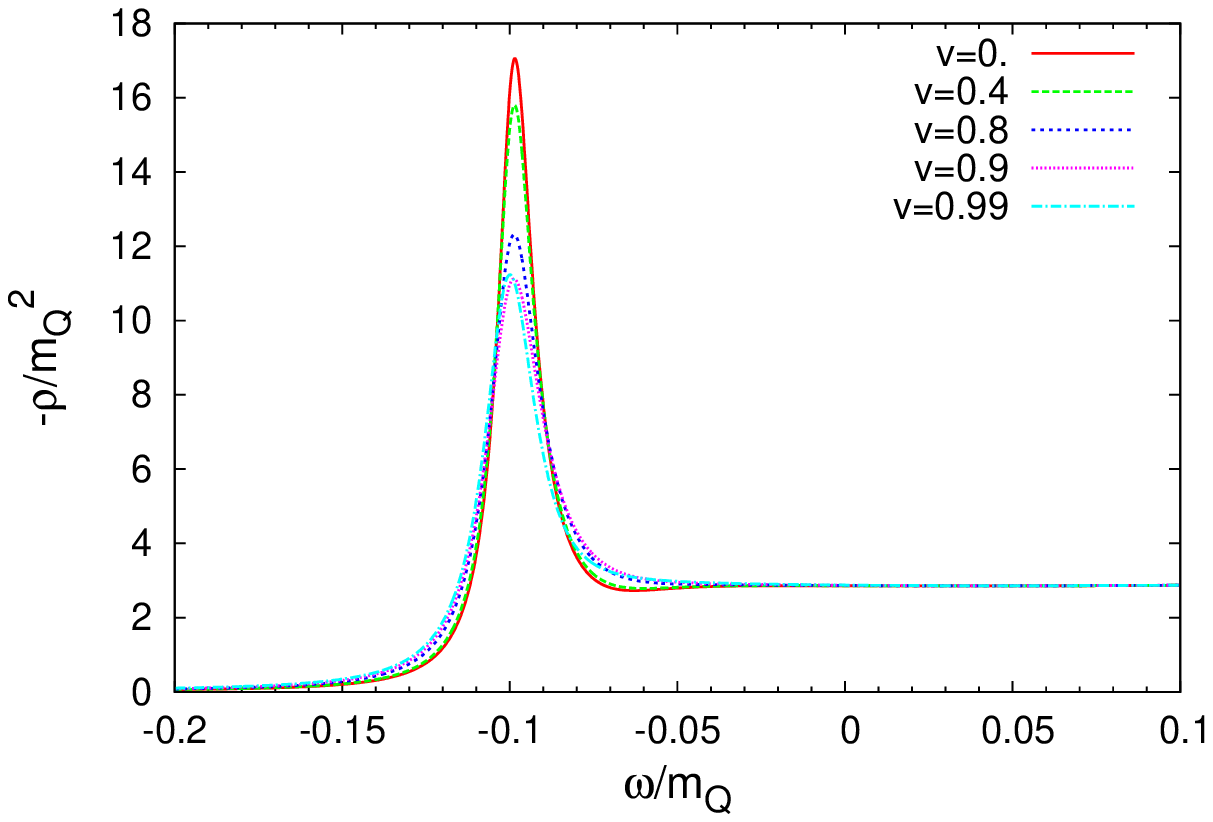}}
\subfigure{
\includegraphics[width=8cm]{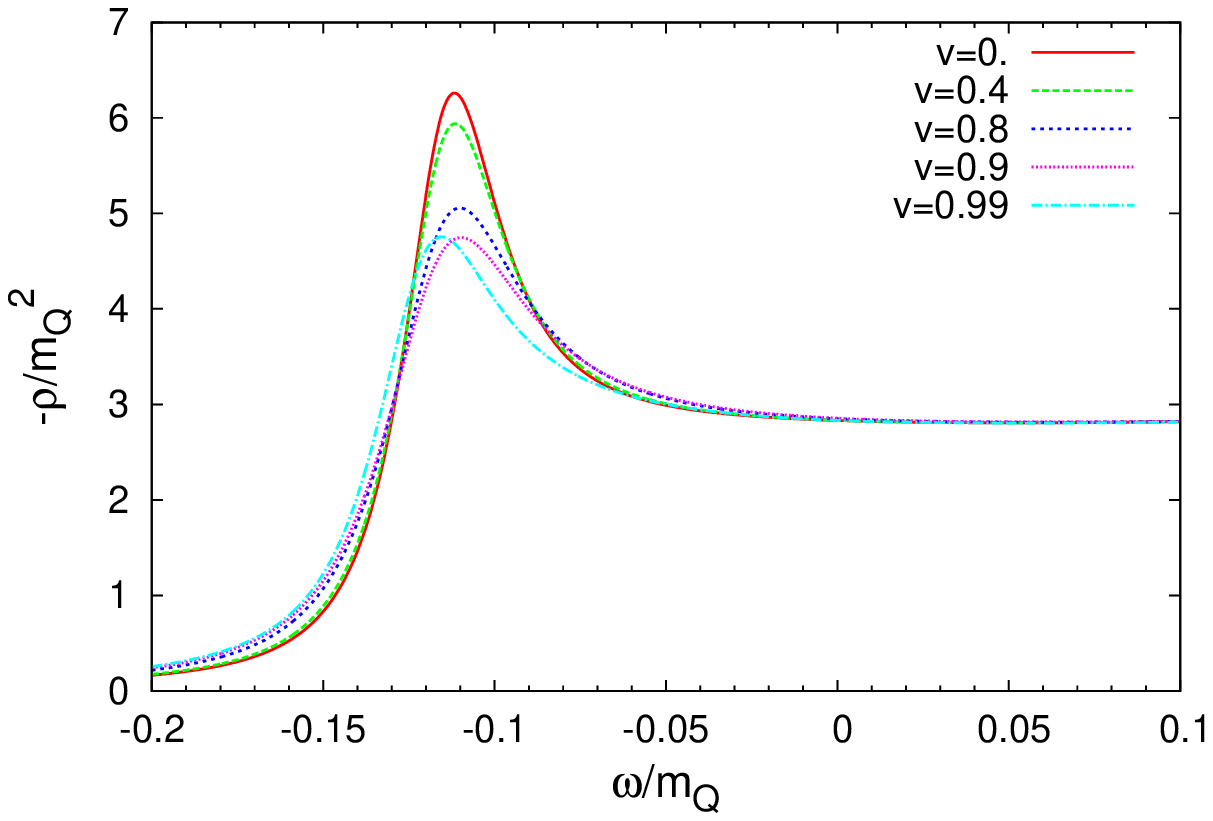}}

 \caption{(Color online) Spectral function of the $\Upsilon(1s)$, divided by $-m_Q^2$, at $T=250$ MeV (left panel) and at $T=400$ MeV (right panel), for various values of the velocity between the bound state and the plasma.}
 \label{fig:SF}
\end{figure}

Then, we consider the effect of a non-vanishing velocity. In Fig.~\ref{fig:SF} we report the spectral functions for the $\Upsilon(1s)$ at $T=250$ MeV (left panel) and at $T=400$ MeV  (right panel) for a few values of the velocity of the plasma.  Comparing Fig.~\ref{fig:SFv0} and Fig.~\ref{fig:SF} it is apparent that  the effect of an increasing  velocity -- at least qualitatively --- is akin  to the effect of an increasing temperature. 
Actually, in this case the position of the peak of the spectral function at $T=250$ MeV   does not seem to change at all. But the main effect  is that with increasing velocity the height of the peak  
decreases  and the corresponding width increases; a behavior that  emulates an increase of the  temperature of the medium. At $T=400$ MeV and $v=0$, the spectral function  has a small peak, which almost disappears at $v \simeq 0.9$. The result at this temperature is qualitatively similar to the one observed in \cite{Ding:2012pt} at their highest temperature, although one has to take into account that we use different reference frames.

It is interesting to notice that the  tendency of the peak to become smaller and wider as the velocity increases, changes when going from $v=0.9$ to $v=0.99$. At $v=0.99$ the $\Upsilon(1s)$ peak  is slightly higher and slightly narrower than  at $v=0.9$, as shown for two different temperatures in  Fig.~\ref{fig:SF}. This behavior is consistent with the result, already discussed in the previous sections, that the bound states become stable at ultra-relativistic velocities.  As in the  particular case considered in the previous section, this behavior can be related to the fact that the effective temperature of the plasma is the one given in Eq.~\eqref{effective-temperature}, and therefore for large $v$ the plasma is almost everywhere cold.
The fact that  $m_D(z,v)$ becomes purely imaginary, implies that the potential ceases to be Yukawa-like and becomes oscillatory, as already observed numerically in Ref.  \cite{Escobedo:2011ie}.

Except for this peculiar behavior
at $v\rightarrow 1$, both the spectral function analysis and the computation of the width through \eqref{eq:gamma1s} show that the width increases as the velocity of the 
plasma increases, as far as  $v \lesssim 0.9$. This is just the opposite of the results of Eq. \eqref{eq:width-swave} in Case I. The reason is that the two results refer to 
different energy regions, which are dominated by different processes. In Case I the thermal width is dominated by gluo-dissociation processes. 
In the present case, the dominant contribution is determined by Landau damping, which is a collisionless process. We shall further comment on this issue in the Sec.~\ref{sec:Conclusion}.

\section{Discussion and Conclusions} \label{sec:Conclusion}

In this section we first discuss how the relative velocity $v$ used throughout is related to measurable quantities in HIC experiments, like the  momentum of the heavy quarkonium state in the lab frame, $P^\mu$, and the local velocity of the QGP, ${\bm w}$, in that frame, and make a rough estimate of the importance of the relative motion in the yields. Next we compare our result with lattice computations, earlier weak-coupling analysis, and AdS/CFT calculations. We close it with the conclusions. 

\begin{figure}
\includegraphics[width=8cm]{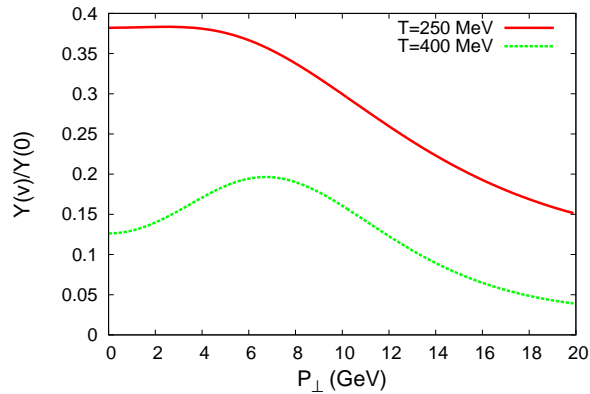}
\caption{(Color online) Ratio of the dilepton yields obtained by Eq.~\eqref{y} for the 1$s$ state versus $P_\bot$ for typical LHC  plasma velocities ($w\sim w_\bot\sim 0.66$). For simplicity we consider  $P_{\parallel}=0$. }
\label{yields}
\end{figure}

The clearest experimental signal of the velocity dependence in the in-medium heavy quarkonium properties should be  in the dilepton yields at fixed rapidity and transverse momentum. 
In order to have an estimate of the effect, we assume that in a central collision the produced medium expands at a constant velocity,   ${\bm w}$, with respect to the lab frame. 
Typical values for ${\bm w}$ quoted in the literature are ${ w}_{\|}\sim 1$ and ${ w}_{\bot}\sim 0.6$ for RHIC and ${ w}_{\bot}\sim 0.66$ for LHC.  We further assume that the system has had enough time to thermalize and that it is isotropic.  A heavy quarkonium produced with a certain $P^\mu$ in that frame,  moves with respect to the plasma with a velocity
\be
{\bm v}=\frac{-P^0{\bm w}+\frac{{\bm P}\cdot {\bm w}}{{\bm w}^2}{\bm w}+\left({\bm P} -\frac{{\bm P}\cdot {\bm w}}{{\bm w}^2}{\bm w}\right) \sqrt{1-{\bm w}^2}}{P^0-{\bm w}\cdot {\bm P}} \; ,
\ee
which is the velocity appearing in the formulas of the previous sections. Notice that for a  given  longitudinal momentum this velocity is not totally fixed, it still depends on  the transverse momentum and on the modulus of the parallel and perpendicular velocities of the plasma in the lab frame, and on the angle $\varphi$ between
${\bm w}_{\bot}$ and ${\bm P}_{\bot}$ in the transverse plain. The modulus of the velocity can be written as
\be
v=\vert {\bm v}\vert =
\sqrt{
1-\frac{\left(1-{\bm w}^2\right)M^2}{M^2-2P^0{\bm w}\cdot {\bm P}+\left( {\bm w}\cdot {\bm P}\right)^2+{\bm P}^2}} \; ,
\label{v}
\ee
where $M$ is the heavy quarkonium mass. Assuming a uniform distribution for the angle $\varphi$,  the modification of the dilepton yields can be estimated by the following
formula
\be
\label{y}
\frac{Y( v)}{Y(v=0)} \sim \frac{1}{2\pi}\int_{0}^{2\pi} d\varphi \, e^{-\left(\Gamma (v)-\Gamma (v=0)\right) \tau } \, ,
\ee
where $\Gamma (v)$ is the velocity dependent decay width calculated in the previous sections evaluated for $v \equiv v(\bm \omega, \bm P)$ given in Eq.~\eqref{v}, $\tau$ is the lifetime of the QGP, about $10$ fm/c for RHIC
or about $15$ fm/c for LHC, and $Y(v)$ stands for the yields. In order to estimate the size of the effect, we display in Fig.~\ref{yields} the results obtained for ${\bm P}_{\|}=0$ and $w\sim w_{\bot} \sim 0.66$ for two different temperatures. 
We plot the ratio between the velocity dependent yield and the yield at $v=0$ as a function of $P_\bot$.  The yield has a non-trivial  dependence on the transverse momentum, which modifies with increasing temperature from a monotonic decreasing behavior at $T \lesssim 250$  MeV to a non-monotonic behavior at higher temperature. 

A number of oversimplifications have been employed in Eq.~\eqref{y}.
We have assumed that the heavy quarkonium decays in the medium and that the medium temperature and the expanding velocity are constant.
These approximations should be reasonable if the decay is much shorter than $\tau$, and since $1/\Gamma(v) \simeq 3$ fm for $\Upsilon(1s)$, this seems the case. In principle  these aspects can be corrected for along the approaches of  \cite{Strickland:2011mw, Strickland:2011aa} or \cite{CasalderreySolana:2012av} (see also \cite{Grandchamp:2005yw,Emerick:2011xu,Sharma:2012dy}).  We have as well assumed  a constant  ${\bm P}$  for the whole evolution, that is we have neglected the damping of the heavy quarkonium, which should be a reasonable approximation because the  drift by the expanding medium is expected to be small. We have also ignored the velocity dependence in the production mechanism and the contribution of the continuum to the yield. 
Eq.~\eqref{y}  is a reasonable approximation if all the above-mentioned corrections factor out in the yield; which might not be the case. However, we postpone to future work a more reliable estimate of this quantity. In any case, we believe that Eq.~\eqref{y}, together with Fig.~\ref{yields}, is enough to pinpoint the importance of the relative velocity between the heavy quarkonium states and the thermal medium.

The spectral function of the 
bottomonium states in a moving thermal bath have been studied with different lattice methods  in \cite{Oktay:2010tf, Aarts:2012ka, Ding:2012pt, Nonaka:2011zz}.
It is important to take into account that while our computation is performed in the heavy quarkonium rest frame, lattice computations are done in the thermal bath rest frame. The spectral function in the heavy quarkonium rest frame $\rho_{\text{HQ}}(k_0)$ where $k_0=p_0-M$ and 
$p^2\sim M^2$, is related to the spectral function in the plasma rest frame $\rho_{\text{plasma}}$ by the following equation
\begin{equation}
\rho_{\text{plasma}}(k_0)=\eta(v)\rho_{\text{HQ}}(k_0/\gamma) \,,
\end{equation}
where $\gamma=1/\sqrt{1-v^2}$ and $\eta(v)$ is a function of the velocity that is not important for the discussion below. If the thermal modifications are a perturbation  $\rho_{\text{HQ}}(k_0)$ in the vicinity of a peak is well approximated by a Breit-Wigner distribution
\begin{equation}
\rho_{\text{HQ}}(k_0)=\frac{A(M)}{(k_0-E)^2+(\Gamma/2)^2} \,,
\end{equation}
hence
\begin{equation}
\rho_{\text{plasma}}(k_0)=\frac{\gamma^2\eta(v)A(M)}{(k_0-\gamma E)^2+(\gamma \Gamma/2)^2}\,.
\end{equation}
This means that even if heavy quarkonium is not modified by the velocity of the plasma in the frame where it is at rest, we would still see a modification of the spectral function in the plasma rest frame. This modification will lead to an increase of the energy where the peak is located and a broadening of the peak. 

In the case of Sec. \ref{sec:CaseI}, we can compare with the lattice results of  \cite{Aarts:2012ka}. Unfortunately, the velocity range explored in \cite{Aarts:2012ka}   was at most of the order of $v=0.2$, for which  no velocity dependent width change has been observed. This null result is compatible with our results taking into account the error in the lattice computations and the fact that their temperatures are not very high. Note that the ans\"atze made in 
\cite{Aarts:2012ka} for the binding energy and the decay width as a function of the velocity, based on the hydrogen atom computation in \cite{Escobedo:2011ie}, holds for $s$-wave states only according to our results for QCD. At zero velocity the lattice results of the same group \cite{Aarts:2010ek,Aarts:2011sm} turned  out to be compatible with our results and with those of \cite{Brambilla:2010vq}.
The results of this section for the decay width also appear to be compatible with the weak coupling calculation of Ref. \cite{Song:2007gm} at leading order (LO), which also shows a decreasing behavior of the decay width with the velocity for small screening masses. At that order, only gluo-dissociation diagrams contribute, like in our case.

In the case of Sec. \ref{sec:particular}, the results only hold if the thermal corrections can be considered as a perturbation. This implies that we can only compare to spectral functions that have approximately a Breit-Wigner form. Because of this we cannot compare with the lattice results in \cite{Ding:2012pt} but we can compare with those in \cite{Nonaka:2011zz}. 
 By analyzing Figs. 4 and 5 in \cite{Nonaka:2011zz} we can obtain approximate values of  the decay width of the $\eta_c$ for several momenta. 
The decay widths we obtain from those figures and the corresponding prediction from Eq.~\eqref{eq:gammav}, which is flavor independent and so holds also for charmonium states, are shown in Table \ref{t:comp}. We observe a similar qualitative behavior, since in both cases the width is a non-monotonic function of the velocity, {\it i.e.} it increases for low values of $v$ and decreases for larger $v$, but the value of $v$ at which it starts to decrease is lower in   \cite{Nonaka:2011zz} than in our case.
However, one has to consider  that a more detailed statistical analysis would be necessary to disentangle
possible MEM artifacts from the actual width in the plots of Ref.~\cite{Nonaka:2011zz}, see \cite{Asakawa:2000tr}
\footnote{We thank Masayuki Asakawa for pointing this out to us.}. Moreover, the temperatures at which this comparison is done may be too close to the deconfinement phase transition for the weak coupling expansion used in our computations to be reliable. The equivalent temperature regime for bottomonium
(\textit{i.e.} higher temperature) would be much safer. The results of this section agree with the weak coupling estimate of the dissociation temperature in Ref. \cite{Dominguez:2008be} and also appear to be compatible with the contributions to the decay width at next-to-leading order (NLO) displayed in Ref. \cite{Song:2007gm}. In the last reference, it was found that the NLO contribution was much larger than LO one \footnote{There are a number of approximations in the NLO calculation of \cite{Song:2007gm}, in particular Pauli blocking is ignored, see \cite{Brambilla:2013dpa} for a recent discussion.}. In our EFT approach this can be easily understood if the system is in the kinematical regime of Sec. \ref{sec:CaseII}, in which the gluo-dissociation processes contributing to their LO are parametrically suppressed. This is also consistent with the arguments and results presented in Ref. \cite{Zhao:2007hh}.  
\begin{table}
\begin{tabular}{|c|c|c|c|}
\hline
$p$ & $v$ & $\Gamma_{\text{plot}}$ (MeV) & $\Gamma_{\text{pred}}$ (MeV) \\
\hline
$0$ & $0$ & $106$ & X \\
$6$ & $0.6$ & $135$ & $132$ \\
$7$ & $0.65$ & $134$ & $139$ \\
$8$ & $0.67$ & $128$ & $142$ \\
\hline
\end{tabular}
\caption{Comparison of the results reported in the figure 4 of \cite{Nonaka:2011zz} for the $\eta_c$ state with the prediction of our Eq.~\eqref{eq:gammav}. The first column $p$ is the momentum in the units used in \cite{Nonaka:2011zz} ($\sim$0.5 GeV). The second column is the velocity of the plasma deduced by looking at figure 5 in the same reference. The third column is the width obtained by assuming that the spectral function can be approximated by a Breit-Wigner distribution and comparing the highest point of the peak with the points where the value is half the maximum. Finally the fourth column is the value of the width predicted by using Eq.~\eqref{eq:gammav}, where, for  $\Gamma_{1}^{s-wave}(v=0)$, we have used the corresponding value in the third column (106 MeV).}
\label{t:comp}
\end{table}

A number of analysis on the velocity dependence of the screening length have been carried out for strong coupling using the AdS/CFT approach  \cite{Liu:2006nn,Chernicoff:2006hi,Caceres:2006ta,Avramis:2006em,Natsuume:2007vc,Ejaz:2007hg,Chernicoff:2012bu} (see \cite{CasalderreySolana:2011us} for a review). It is not straightforward to compare these results to ours, as they do not obtain an imaginary part in the potential. This would be a first important difference. Furthermore, in momentum space, what plays the role of the screening mass for us is the complex, angle and velocity dependent, Debye mass $m_D(v,\theta)$, see Fig.\ref{fig:debye-mass} , which translates into a non-trivial potential in coordinate space for which no simple analytical form has been found. Hence we cannot make further statements on this respect than those already made in Ref. \cite{Escobedo:2011ie}\footnote{In formula (92) of  \cite{Escobedo:2011ie}, $m_D(v,\theta)$ should read $\vert m_D(v,\theta)\vert$.}. However, we can certainly 
compare with the two AdS/CFT calculations of the heavy quarkonium
spectral function at non-vanishing velocity we are aware of \cite{Myers:2008cj,Fujita:2009wc}. These spectral functions qualitatively agree with ours in Case II at moderate velocities, in the sense that the bound state peaks become smaller and wider as the velocity increases. 
 Let us finally remark that we observe in the ultrarelativistic limit an oscillatory behavior of the potential rather than an exponential damping, that would lead to the stabilization of the bound states, which is not observed in the AdS/CFT approach. 

In summary, we have analyzed heavy quarkonium states moving in a weakly coupled QCD plasma. In the Case I, corresponding to  the 
hierarchy $m_Q \gg 1/r\gg T \gg E\gg m_D$, we have found that the thermal decay width decreases as the velocity increases, like in  QED \cite{Escobedo:2011ie}. The decay width is in this case dominated by gluo-dissociation processes \cite{Brambilla:2011sg}. However, unlike in QED, the thermal energy shift becomes velocity dependent, except for the $s$-wave states. In the Case II, corresponding to  the 
hierarchy $m_Q \gg T \gg 1/r \,, m_D \gg E$, we have found a different behavior for the decay width, namely it increases as the velocity increases, except for ultra-relativistic velocities for which it starts decreasing again. This non-trivial behavior was overlooked in Ref.\cite{Escobedo:2011ie}. The decay width is in this case dominated by the Landau damping.
Putting all together, we conclude that the decay width depends in a nontrivial way on the temperature and on the velocity, which complicates the interpretation of  HIC experimental data, as we tried to illustrate by Fig.~\ref{yields}. Our results are consistent and in qualitative and semi-quantitative agreement with the few available lattice data \cite{Aarts:2012ka,Nonaka:2011zz},
and  also appear to be compatible with the weak coupling analysis of refs.  \cite{Song:2007gm,Dominguez:2008be}.

\bigskip

{\bf Acknowledgments}

\bigskip

We thank Mikko Laine for providing us with the codes used in Refs. \cite{Laine:2007gj,Burnier:2007qm} and for discussion.  We also thank Ralf Rapp for bringing to our attention a number of references. JS has been supported by the HadronPhysics3 project,
FP7-Infrastructures-2011-1 Grant agreement 283286 (EU), the CPAN  CSD2007-00042 Consolider-Ingenio 2010 program (Spain), the FPA2010-16963 project (Spain) and the 2009SGR502 CUR grant (Catalonia).
FG has been supported in part by the Italian MIUR Prin 2009. MAE was supported by the DFG grant BR4058/1-1 and by the European Community under the FP7 programme HadronPhysics3.

\appendix
\section{General Framework}
\label{appendix-gen}
In this appendix we briefly review the general framework  used  to take into account the effect of a moving thermal medium. A more detailed discussion can be found in \cite{Escobedo:2011ie}. 

\begin{center}
\begin{figure}[h!]
\includegraphics[width=8cm]{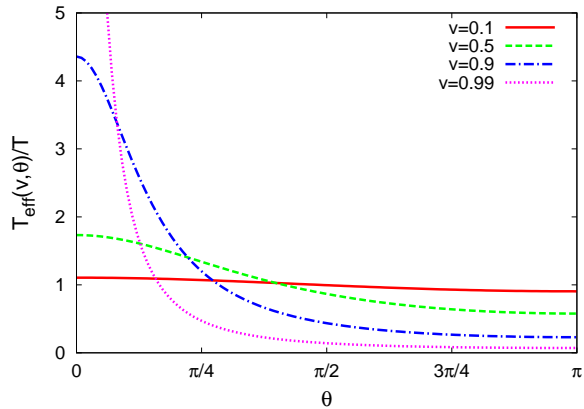}
 \caption{(Color online) Effective temperature divided by $T$, as a  function of the angle between the vectors ${\bm k}$ and ${\bm v}$, for $v=0.1, 0.5, 0.9, 0.99$.   }
 \label{fig:effectiveT}
\end{figure}
\end{center}

We shall assume that the plasma (or black-body radiation) is in  thermal equilibrium at a temperature $T$.  Since we are 
considering the reference frame in which the plasma is  moving with a velocity $\bm v$, the particle distribution functions are given by
\be\label{boosted-distributions}
f_{F,B} (\beta^\mu k_\mu ) =\frac{1}{e^{|\beta^\mu k_\mu|}\pm1},
\ee
where the plus (minus) sign refers to fermions (bosons). In the reference frame where the thermal bath is at rest
$\beta^\mu k_\mu=\frac{k_0}{T}$, while  in a frame where the plasma moves with a velocity $\bm v$ we have that
\be 
\beta^\mu = \frac{\gamma}{T}(1,{\bm v})= \frac{u^\mu}{T} \,, 
\ee
where $\gamma= 1/\sqrt{1-v^2}$ is the Lorentz factor; the latter frame
 has been successfully used in the past, for example in~\cite{Weldon:1982aq}.
Studying a bound state in a moving thermal bath is akin to study a bound state in  non-equilibrium field theory~\cite{Carrington:1997sq};
in that case the Bose-Einstein or Fermi-Dirac distribution functions are substituted by a general distribution, 
which in our case 
will be the boosted Bose-Einstein or Fermi-Dirac distribution functions reported in  Eq.~(\ref{boosted-distributions}). For  a thermal medium formed of massless particles, taking into account that in non-equilibrium field theory the collective behavior always enters through on-shell particles or antiparticles,  we have (in the case of particles) that
\be
\label{dist}
\beta^\mu k_\mu=k\frac{1- v \cos{\theta}}{T\sqrt{1-v^2}}\,,
\ee
where $k=\vert {\bm k}\vert$ and $\theta$ is the angle between $\bm{k}$ and $\bm{v}$. The distribution functions in
Eq.~(\ref{boosted-distributions}) can now be written as 
\be
\label{boosted-2} f_{F,B}(k,T, \theta, v)=\frac{1}{e^{k/T_{\rm eff}(\theta,v)}\pm 1}\,,
\ee
where we have defined the effective temperature
\be\label{effective-temperature2}
T_\textit{eff}(\theta,v)=\frac{T\sqrt{1-v^2}}{1-v\cos{\theta}}\,,
\ee
which is plotted in Fig.~\ref{fig:effectiveT} for few values of $v$.  Fig.~\ref{fig:effectiveT}  helps to clear away the misconception that a bound state moving with  non-vanishing velocity in a thermal bath feels a higher temperature. Indeed, the effective temperature is in most of the directions smaller than $T$; for $v \sim 1$ we find that $T_\textit{eff}(\theta,v)>T$ only for $0< \theta < \sqrt{2}(1-v^2)^{1/4}$. Intuitively, the dependence of the effective temperature on $v$ and $\theta$ can be understood as a Doppler effect.

 While at $v=0$ it is clear that the thermal medium introduces a new scale $T$ in the problem, it is not clear {\it a priori} how many scales a moving thermal medium introduces. This can be understood by using light-cone coordinates.  We choose $\bm v$  in the $z$ direction and define
\be k_+=k_0+k_3 \qquad {\rm and}  \qquad k_-=k_0-k_3 \,.
\label{k+-}
\ee
Then, we have  that
\begin{equation}
\beta^\mu k_\mu=\frac{1}{2}\left(\frac{k_+}{T_+}+\frac{k_-}{T_-}\right) \,,
\label{exp}
\end{equation}
where
\be\label{temperatures}
T_+=T\sqrt{\frac{1+v}{1-v}} \qquad {\rm and} \qquad T_-=T\sqrt{\frac{1-v}{1+v}} \,.
\ee
Therefore, in light-cone coordinates, it becomes explicit that the distribution function actually depends on two scales, $T_+$ and $T_-$. Obviously, for any  value of $v$, one has  that $T_+ \ge T \ge T_-$, and moreover $T_+$ corresponds to the highest temperature measurable by the observer, while $T_-$ corresponds to the lowest temperature measurable by the observer. In this work we consider always that $T_+$ and $T_-$ are of the same order of magnitude. Even though this is not so for very large velocities, in all the cases considered in \cite{Escobedo:2011ie} we found the results obtained assuming $T_+\sim T_-$ were indeed correct even for $v\rightarrow 1$.

\section{Numerical checks}\label{appendix-osm}
The operator-splitting is a powerful method for solving partial differential equations \cite{Bagrinovskii}. The  idea is to separate a complex differential equation in various simpler equations and to solve them with  a discretization  method. For illustrative purposes  we compare the results obtained for a hydrogen-like atom with two different potentials, namely the Yukawa potential
\be \label{eq:yukawa}
V_{\rm Y}({\bm x})=-\frac{\alpha}{\sqrt{x^2+y^2+z^2}} e^{- m_{\rm D} \sqrt{x^2+y^2+z^2} }
\,,\ee
and  the potential reported in \cite{Escobedo:2011ie} for vanishing velocity of the thermal medium. These analyses could be done employing a numerical code with spherical symmetry, however using cylindrical coordinates 
allows us to check the numerical procedure employed in the general case of non-vanishing velocity. Indeed, the present discussion  can be generalized to any potential with cylindrical symmetry.

The Schr\"odinger equation in cylindrical coordinates, $(z,r,\varphi)$, for $s$-wave states is given by (for simplicity we set $m_e=1, \alpha=1$) 
\be
i \frac{\partial u(t,r,z)}{\partial t}=-\left(\frac{1}{2}\frac{\partial^2}{\partial r^2}+\frac{1}{2}\frac{\partial^2 }{\partial z^2}-\frac{1}{2r}\frac{\partial }{\partial r}+ \frac{1}{2r^2} - V(r,z) \right) u(t, r,z)\,,
\ee  
being $u(t,r,z)=r\psi(t,r,z)$; in order to compute the spectral function we use the initial condition $u(0,r,z)=-r\delta^2(r)\delta(z)$ and boundary condition $u(t,0,0) = 0$.
We separate the Hamiltonian in one term containing the potential and derivatives with respect to $z$ and    a second term with only derivatives with respect to $r$: 
\bea
H_1&=&-\frac{1}{2}\frac{\partial^2 }{\partial z^2}+ V(r,z)\,, \\
H_2 &=&-\frac{1}{2}\frac{\partial^2}{\partial r^2}+\frac{1}{2r}\frac{\partial }{\partial r}- \frac{1}{2r^2} \,,
\eea
and we solve the corresponding  Schr\"odinger equations numerically using the Crank-Nicolson method \cite{crank-nicolson}, meaning that the equations are discretized as follows,
\bea
\left(1+\frac{i}{2} H_1 a_t \right)u^{n+1/2}&=& \left(1-\frac{i}{2} H_1 a_t \right)u^{n}\,,\label{eq:systemH1}\\
\left(1+\frac{i}{2} H_2 a_t \right)u^{n+1}&=&\left(1-\frac{i}{2} H_2 a_t \right)u^{n+1/2}\,,
\label{eq:systemH2}
\eea
where $a_t$ is the temporal lattice spacing and  $n$ indicates the discretized time step. The equations are solved recursively: At the $n$-th  step the wave-function evolves from time $n a_t$ to $(n+1/2) a_t$ according to Eq.~\eqref{eq:systemH1}.  In the next step from $(n+1/2) a_t$ to $(n+1)a_t$ according to Eq.~\eqref{eq:systemH2}. In  Eqs.~\eqref{eq:systemH1} and \eqref{eq:systemH2} also the space coordinates are discretized  $z=l a_s$ and $r=j a_s$, where $a_s$ is the spatial lattice spacing and $l,j$ are integers.

The discretized  initial condition  reads  
\bea
r\delta^2(r)\delta(z) &=& \frac{r}{8\pi^3}\int d^2 {\bm p_r} \, e^{i 
{\bm p_r \cdot \bm r}}\int_{-\infty}^{\infty} dp_z \, e^{ip_z z} \nonumber\\
&\to & \frac{ja_s}{4\pi^2} \int_0^{\pi/a_s} dp_r\, \frac{2}{a_s}\, J_1(p_r 
a_s)\, J_0(jp_r a_s) \int_{-\pi/a_s}^{\pi/a_s} dp_z \, e^{ip_z l a_s}  \nonumber\\
&=&  \frac{j}{\pi\, a_s^2} \delta_{l0} \int_0^{\pi} du\,  J_1(u)\, J_0(j\, u) \,,
\eea
 where the   Bessel functions of the first kind, $J_\alpha$, are used instead of the trigonometric function of \cite{Laine:2007gj} to improve the convergence. Once the discretized wave functions are obtained,   the corresponding spectral function can be calculated by means of Eq.~\eqref{eq:specfun}.

We consider first the Yukawa potential. The binding energy of the various states is known with great accuracy, see \textit{e.g.} \cite{harris, rogers, Vrscay}, and it is also known that sequential dissociation of the bound states  takes place 
with increasing values of $\lambda= m_D  a_0  $, where $a_0$ is the Bohr radius. 
\begin{figure}[ht!]
\begin{center}
\subfigure{
 \includegraphics[width=8cm]{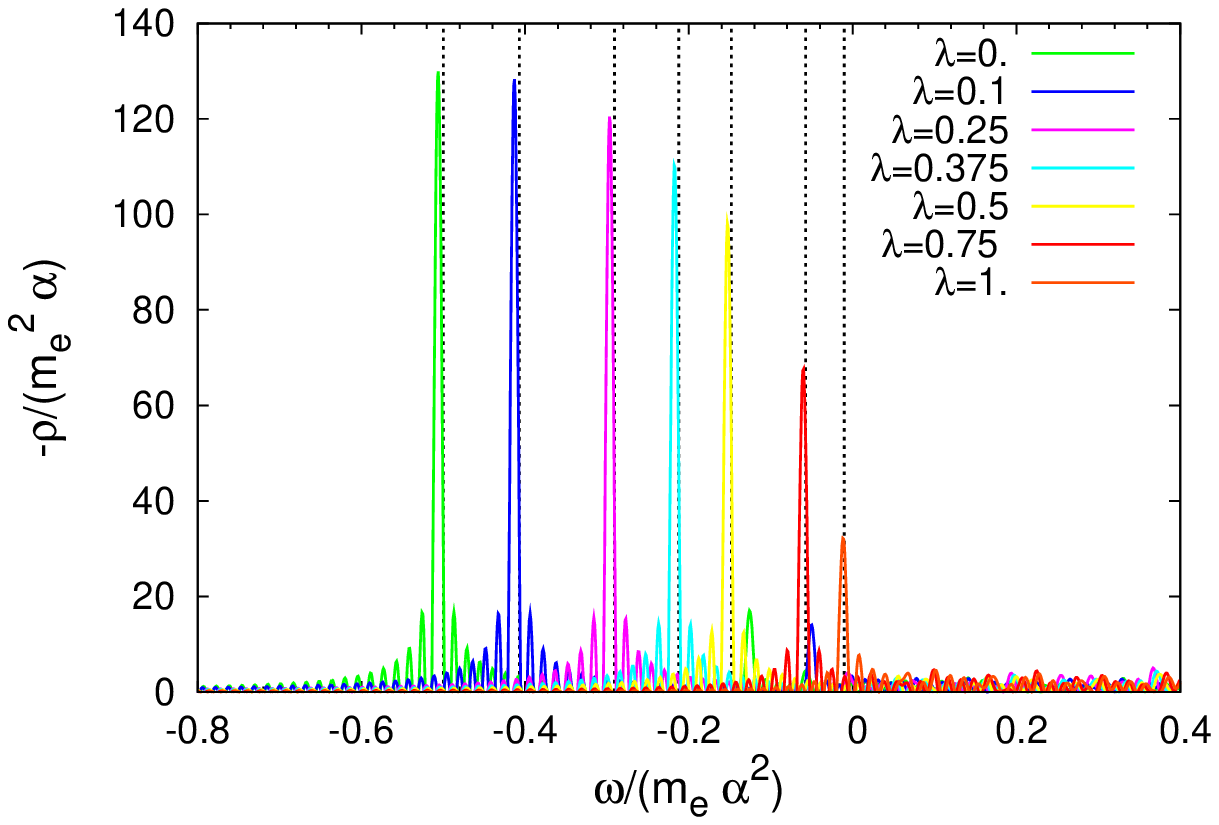}
 }
\subfigure{
\includegraphics[width=8cm]{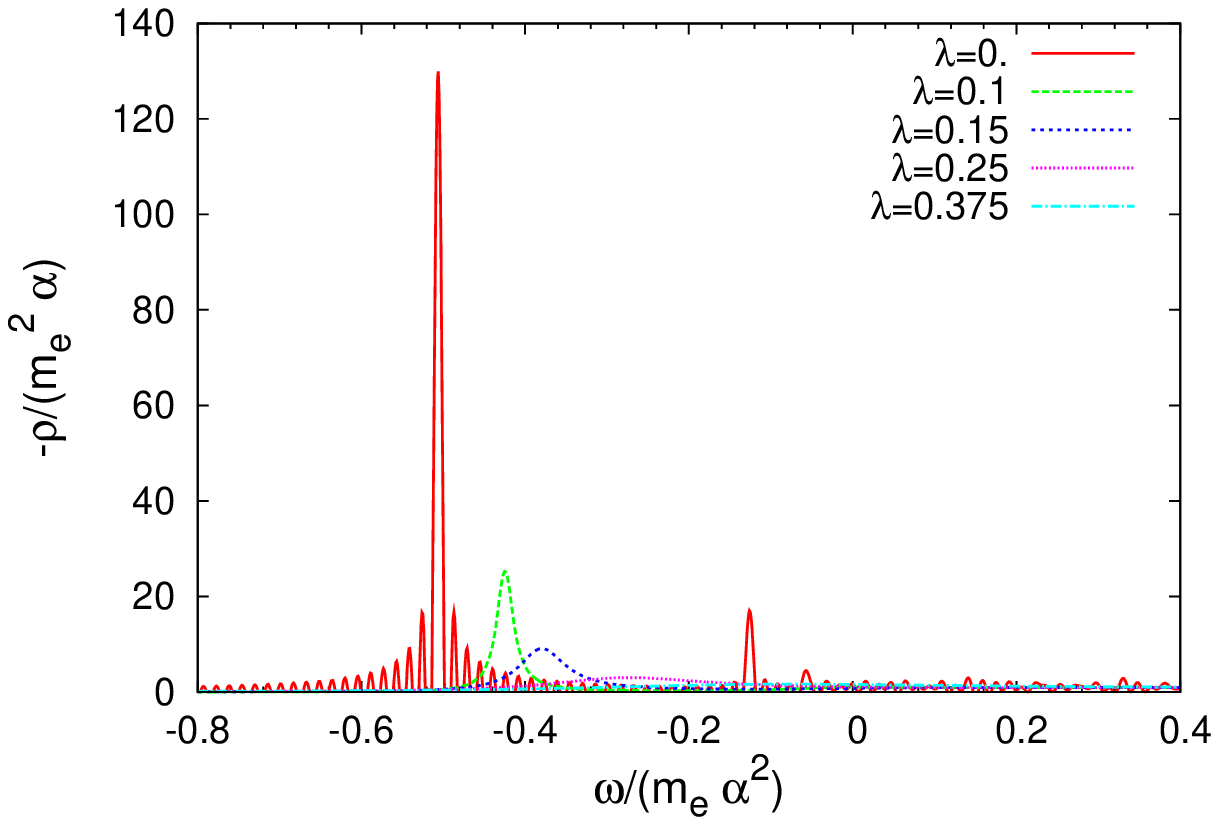}}
\caption{(Color online) Left panel: Spectral functions obtained for the hydrogen-like atom with the Yukawa potential \eqref{eq:yukawa} for various values of $\lambda=m_D/(m_e\alpha)$. The numerical algorithm slightly overestimates the binding energies obtained in \cite{Vrscay}, corresponding to the  vertical dotted lines.
Right panel: Spectral functions obtained with the potential reported in \cite{Escobedo:2011ie} for vanishing velocity. }
\label{fig:Hydrogen-Yukawa}
\end{center}
\end{figure}
On the left panel of Fig.~\ref{fig:Hydrogen-Yukawa} we report the spectral functions  obtained with the splitting method for various values of $\lambda$; the vertical lines correspond to the numerical values of the binding energy obtained in \cite{Vrscay} for the $1 s$ state.   As expected, at $m_D=0$ (green line) we find a peak at $\omega \simeq -0.5 m_e\alpha^2$,  corresponding to the $1s$ state of the standard hydrogen atom. The $2s$ state at $\omega \simeq -0.125 m_e\alpha^2$ is also visible, but the corresponding height  is  suppressed. 
In principle, at any bound state there should exist a corresponding Dirac delta-function, however the discretization procedure can only lead to a finite peak. The height of the peak is proportional to the  field strength, that is to the strength of the corresponding interaction channel, which explains the fact that the $2s$ peak is smaller than the $1 s$ peak.

With increasing values of the Debye mass  the peak of the $1s$ state  moves to higher values of  energy, meaning that the corresponding binding energy decreases. Although the peak height decreases, for the same reason explained above,  note that no appreciable broadening of the spectral function appears,   meaning that the numerical procedure does not produce a fictitious increase of the width. Indeed, in this simple model (and in any model with a real potential) the dissociation happens when the peak of the spectral function approaches zero. Our numerical results indicate that  the $1s$ state of the Yukawa potential dissociates at $\lambda \simeq 1.2 $,  in good agreement with the  numerical results of  \cite{rogers, Vrscay}.

Regarding the $2 s$ state, at $\lambda =0.1$ (blue line) it is still visible, with binding energy  $\omega \simeq -0.05 m_e\alpha^2$ (in good agreement with the results of \cite{rogers, Vrscay}), but for larger values of $\lambda$ the $2s$ state is no more visible, although it is known that it only dissociates  at $\lambda \simeq 0.31$, see  \cite{rogers, Vrscay}.   The reason, as explained above, is that the corresponding peak is very small and cannot be identified with the used numerical accuracy.

In summary, from the analysis of the Yukawa potential, we conclude that the algorithm correctly reproduces the binding energy of the $1s$ state at any value of the Debye mass, but the decrease of the peak height observed in the left panel of Fig.~\ref{fig:Hydrogen-Yukawa} for increasing values of $m_D$ is an artifact due to the combined effect of the numerical discretization and of the  reduction of the strength of the  interaction channel. Remarkably, the algorithm does not produce a fictitious width. The analysis of the dissociation of the excited states for this model with the present method is problematic, because of the reduction of the peak height.

On the right panel of Fig.~\ref{fig:Hydrogen-Yukawa}, we show  the spectral functions obtained with the potential reported in \cite{Escobedo:2011ie} considering vanishing velocity. This potential has an imaginary component for any non-vanishing value of the temperature. At $T=0$, \textit{i.e.} for $\lambda=0$,  the potential is real and Coulombic, and the  standard  peaks  of the hydrogen atom for the $1s$ state  at $\omega\simeq -0.5 m_e\alpha^2$, the  $2s$ state at  $\omega\simeq -0.125 m_e\alpha^2$ and the $3s$ state at  $\omega\simeq -0.05 m_e\alpha^2$ are reproduced with a  good accuracy. As before, the peaks of states with high principal quantum number are suppressed.  Increasing the temperature, the binding energy of the $1s$ state decreases, but the corresponding spectral function  not only moves to  higher energies, it also becomes wider.  From the insight gained in the  analysis of the  Yukawa potential,  we conclude that  the broadening of the peak is due to the  imaginary part of the potential 
and not to the numerical procedure. Moreover, improving the discretization procedure, we checked that the reduction of the peak height of the $1s$ state (at nonvanishing temperature) is not an artifact, but it is  instead a genuine effect related to the imaginary part of the potential. The reason is that with a finite imaginary potential, the spectral functions are not delta-functions, but smoother functions which can be resolved with the used  discretization method.

\bibliographystyle{ieeetr}
\bibliography{biblio-hq}

\begin{thebibliography}{10}

\bibitem{Matsui:1986dk}
T.~Matsui and H.~Satz, ``{J/psi Suppression by Quark-Gluon Plasma Formation},''
  {\em Phys.Lett.}, vol.~B178, p.~416, 1986.

\bibitem{Karsch:1987pv}
F.~Karsch, M.~Mehr, and H.~Satz, ``{Color Screening and Deconfinement for Bound
  States of Heavy Quarks},'' {\em Z.Phys.}, vol.~C37, p.~617, 1988.

\bibitem{Rapp:2008tf}
R.~Rapp, D.~Blaschke, and P.~Crochet, ``{Charmonium and bottomonium production
  in heavy-ion collisions},'' {\em Prog.Part.Nucl.Phys.}, vol.~65,
  pp.~209--266, 2010.

\bibitem{Mocsy:2013syh}
A.~Mocsy, P.~Petreczky, and M.~Strickland, ``{Quarkonia in the Quark Gluon
  Plasma},'' {\em Int.J.Mod.Phys.}, vol.~A28, p.~1340012, 2013.

\bibitem{Abelev:2010am}
B.~Abelev {\em et~al.}, ``{$\Upsilon$ cross section in $p+p$ collisions at
  $\sqrt(s) = 200$ GeV},'' {\em Phys.Rev.}, vol.~D82, p.~012004, 2010.

\bibitem{Abreu:1997jh}
M.~Abreu {\em et~al.}, ``{Anomalous J / psi suppression in Pb - Pb interactions
  at 158 GeV/c per nucleon},'' {\em Phys.Lett.}, vol.~B410, pp.~337--343, 1997.

\bibitem{Adare:2011yf}
A.~Adare {\em et~al.}, ``{$J/\psi$ suppression at forward rapidity in Au+Au
  collisions at $\sqrt{s_{NN}}=200$ GeV},'' {\em Phys.Rev.}, vol.~C84,
  p.~054912, 2011.

\bibitem{Strickland:2012as}
M.~Strickland, ``{Bottomonia in the Quark Gluon Plasma},'' 2012.

\bibitem{Silvestre:2011ei}
C.~Silvestre, ``{Quarkonia Measurements by the CMS Experiment in pp and PbPb
  Collisions},'' {\em J.Phys.}, vol.~G38, p.~124033, 2011.

\bibitem{Chatrchyan:2011pe}
S.~Chatrchyan {\em et~al.}, ``{Indications of suppression of excited $\Upsilon$
  states in PbPb collisions at $\sqrt{S_{NN}}$ = 2.76 TeV},'' {\em
  Phys.Rev.Lett.}, vol.~107, p.~052302, 2011.

\bibitem{Chatrchyan:2012fr}
S.~Chatrchyan {\em et~al.}, ``{Observation of sequential Upsilon suppression in
  PbPb collisions},'' {\em Phys.Rev.Lett.}, vol.~109, p.~222301, 2012.

\bibitem{Reed:2011fr}
R.~Reed, ``{Measuring the Upsilon Nuclear Modification Factor at STAR},'' {\em
  J.Phys.}, vol.~G38, p.~124185, 2011.

\bibitem{Escobedo:2008sy}
M.~A. Escobedo and J.~Soto, ``{Non-relativistic bound states at finite
  temperature (I): The Hydrogen atom},'' {\em Phys.Rev.}, vol.~A78, p.~032520,
  2008.

\bibitem{Brambilla:2008cx}
N.~Brambilla, J.~Ghiglieri, A.~Vairo, and P.~Petreczky, ``{Static
  quark-antiquark pairs at finite temperature},'' {\em Phys.Rev.}, vol.~D78,
  p.~014017, 2008.

\bibitem{Brambilla:2010vq}
N.~Brambilla, M.~A. Escobedo, J.~Ghiglieri, J.~Soto, and A.~Vairo, ``{Heavy
  Quarkonium in a weakly-coupled quark-gluon plasma below the melting
  temperature},'' {\em JHEP}, vol.~1009, p.~038, 2010.

\bibitem{Escobedo:2010tu}
M.~A. Escobedo and J.~Soto, ``{Non-relativistic bound states at finite
  temperature (II): the muonic hydrogen},'' {\em Phys.Rev.}, vol.~A82,
  p.~042506, 2010.

\bibitem{Caswell:1985ui}
W.~Caswell and G.~Lepage, ``{Effective Lagrangians for Bound State Problems in
  QED, QCD, and Other Field Theories},'' {\em Phys.Lett.}, vol.~B167, p.~437,
  1986.

\bibitem{Brambilla:2004jw}
N.~Brambilla, A.~Pineda, J.~Soto, and A.~Vairo, ``{Effective field theories for
  heavy quarkonium},'' {\em Rev.Mod.Phys.}, vol.~77, p.~1423, 2005.

\bibitem{Pineda:2011dg}
A.~Pineda, ``{Review of Heavy Quarkonium at weak coupling},'' {\em
  Prog.Part.Nucl.Phys.}, vol.~67, pp.~735--785, 2012.

\bibitem{Escobedo:2011ie}
M.~A. Escobedo, J.~Soto, and M.~Mannarelli, ``{Non-relativistic bound states in
  a moving thermal bath},'' {\em Phys.Rev.}, vol.~D84, p.~016008, 2011.

\bibitem{Laine:2006ns}
M.~Laine, O.~Philipsen, P.~Romatschke, and M.~Tassler, ``{Real-time static
  potential in hot QCD},'' {\em JHEP}, vol.~0703, p.~054, 2007.

\bibitem{Laine:2008cf}
M.~Laine, ``{How to compute the thermal quarkonium spectral function from first
  principles?},'' {\em Nucl.Phys.}, vol.~A820, pp.~25C--32C, 2009.

\bibitem{Brambilla:2011sg}
N.~Brambilla, M.~A. Escobedo, J.~Ghiglieri, and A.~Vairo, ``{Thermal width and
  gluo-dissociation of quarkonium in pNRQCD},'' {\em JHEP}, vol.~1112, p.~116,
  2011.

\bibitem{Brambilla:2013dpa}
N.~Brambilla, M.~A. Escobedo, J.~Ghiglieri, and A.~Vairo, ``{Thermal width and
  quarkonium dissociation by inelastic parton scattering},'' 2013.

\bibitem{Grandchamp:2001pf}
L.~Grandchamp and R.~Rapp, ``{Thermal versus direct J / Psi production in
  ultrarelativistic heavy ion collisions},'' {\em Phys.Lett.}, vol.~B523,
  pp.~60--66, 2001.

\bibitem{Petreczky:2010tk}
P.~Petreczky, C.~Miao, and A.~Mocsy, ``{Quarkonium spectral functions with
  complex potential},'' {\em Nucl.Phys.}, vol.~A855, pp.~125--132, 2011.

\bibitem{Mannarelli:2005pz}
M.~Mannarelli and R.~Rapp, ``{Hadronic modes and quark properties in the
  quark-gluon plasma},'' {\em Phys.Rev.}, vol.~C72, p.~064905, 2005.

\bibitem{Riek:2010fk}
F.~Riek and R.~Rapp, ``{Quarkonia and Heavy-Quark Relaxation Times in the
  Quark-Gluon Plasma},'' {\em Phys.Rev.}, vol.~C82, p.~035201, 2010.

\bibitem{Rothkopf:2011db}
A.~Rothkopf, T.~Hatsuda, and S.~Sasaki, ``{Complex Heavy-Quark Potential at
  Finite Temperature from Lattice QCD},'' {\em Phys.Rev.Lett.}, vol.~108,
  p.~162001, 2012.

\bibitem{Burnier:2012az}
Y.~Burnier and A.~Rothkopf, ``{Disentangling the timescales behind the
  non-perturbative heavy quark potential},'' {\em Phys.Rev.}, vol.~D86,
  p.~051503, 2012.

\bibitem{Akamatsu:2011se}
Y.~Akamatsu and A.~Rothkopf, ``{Stochastic potential and quantum decoherence of
  heavy quarkonium in the quark-gluon plasma},'' {\em Phys.Rev.}, vol.~D85,
  p.~105011, 2012.

\bibitem{Akamatsu:2012vt}
Y.~Akamatsu, ``{Real-time quantum dynamics of heavy quark systems at high
  temperature},'' 2012.

\bibitem{Adare:2006nq}
A.~Adare {\em et~al.}, ``{Energy Loss and Flow of Heavy Quarks in Au+Au
  Collisions at s(NN)**(1/2) = 200-GeV},'' {\em Phys.Rev.Lett.}, vol.~98,
  p.~172301, 2007.

\bibitem{Awes:2008qi}
T.~C. Awes, ``{Highlights from PHENIX - II},'' {\em J.Phys.}, vol.~G35,
  p.~104007, 2008.

\bibitem{vanHees:2007me}
H.~van Hees, M.~Mannarelli, V.~Greco, and R.~Rapp, ``{Nonperturbative
  heavy-quark diffusion in the quark-gluon plasma},'' {\em Phys.Rev.Lett.},
  vol.~100, p.~192301, 2008.

\bibitem{Chu:1988wh}
M.~Chu and T.~Matsui, ``{DYNAMIC DEBYE SCREENING FOR A HEAVY ANTI-QUARK PAIR
  TRAVERSING A QUARK - GLUON PLASMA},'' {\em Phys.Rev.}, vol.~D39, p.~1892,
  1989.

\bibitem{Pineda:1997bj}
A.~Pineda and J.~Soto, ``{Effective field theory for ultrasoft momenta in NRQCD
  and NRQED},'' {\em Nucl.Phys.Proc.Suppl.}, vol.~64, pp.~428--432, 1998.

\bibitem{Brambilla:1999xf}
N.~Brambilla, A.~Pineda, J.~Soto, and A.~Vairo, ``{Potential NRQCD: An
  Effective theory for heavy quarkonium},'' {\em Nucl.Phys.}, vol.~B566,
  p.~275, 2000.

\bibitem{Weldon:1982aq}
H.~A. Weldon, ``{Covariant Calculations at Finite Temperature: The Relativistic
  Plasma},'' {\em Phys.Rev.}, vol.~D26, p.~1394, 1982.

\bibitem{Brambilla:2011mk}
N.~Brambilla, M.~A. Escobedo, J.~Ghiglieri, and A.~Vairo, ``{The spin-orbit
  potential and Poincar\'e invariance in finite temperature pNRQCD},'' {\em
  JHEP}, vol.~1107, p.~096, 2011.

\bibitem{McLerran:1984ay}
L.~D. McLerran and T.~Toimela, ``{Photon and Dilepton Emission from the Quark -
  Gluon Plasma: Some General Considerations},'' {\em Phys.Rev.}, vol.~D31,
  p.~545, 1985.

\bibitem{Weldon:1990iw}
H.~Weldon, ``{Reformulation of finite temperature dilepton production},'' {\em
  Phys.Rev.}, vol.~D42, pp.~2384--2387, 1990.

\bibitem{Pineda:1998id}
A.~Pineda, ``{Heavy quarkonium and nonrelativistic effective field theories},''
  1998.

\bibitem{Beneke:1997zp}
M.~Beneke and V.~A. Smirnov, ``{Asymptotic expansion of Feynman integrals near
  threshold},'' {\em Nucl.Phys.}, vol.~B522, pp.~321--344, 1998.

\bibitem{Mannarelli:2007gi}
M.~Mannarelli and C.~Manuel, ``{Chromohydrodynamical instabilities induced by
  relativistic jets},'' {\em Phys.Rev.}, vol.~D76, p.~094007, 2007.

\bibitem{Mannarelli:2007hj}
M.~Mannarelli and C.~Manuel, ``{Jet-induced gauge field instabilities in the
  quark-gluon plasma: A Kinetic theory approach},'' {\em Phys.Rev.}, vol.~D77,
  p.~054018, 2008.

\bibitem{Laine:2007gj}
M.~Laine, ``{A Resummed perturbative estimate for the quarkonium spectral
  function in hot QCD},'' {\em JHEP}, vol.~0705, p.~028, 2007.

\bibitem{Burnier:2007qm}
Y.~Burnier, M.~Laine, and M.~Vepsalainen, ``{Heavy quarkonium in any channel in
  resummed hot QCD},'' {\em JHEP}, vol.~0801, p.~043, 2008.

\bibitem{Kajantie:1997tt}
K.~Kajantie, M.~Laine, K.~Rummukainen, and M.~E. Shaposhnikov, ``{3-D SU(N) +
  adjoint Higgs theory and finite temperature QCD},'' {\em Nucl.Phys.},
  vol.~B503, pp.~357--384, 1997.

\bibitem{Ding:2012pt}
H.-T. Ding, ``{Momentum dependences of charmonium properties from lattice
  QCD},'' 2012.

\bibitem{Strickland:2011mw}
M.~Strickland, ``{Thermal $\upsilon_{1s}$ and $chi_b1$ suppression in
  $\sqrt{s_{NN}}=2.76$ TeV Pb-Pb collisions at the LHC},'' {\em
  Phys.Rev.Lett.}, vol.~107, p.~132301, 2011.

\bibitem{Strickland:2011aa}
M.~Strickland and D.~Bazow, ``{Thermal Bottomonium Suppression at RHIC and
  LHC},'' {\em Nucl.Phys.}, vol.~A879, pp.~25--58, 2012.

\bibitem{CasalderreySolana:2012av}
J.~Casalderrey-Solana, ``{Dynamical Quarkonia Suppression in a QGP-Brick},''
  2012.

\bibitem{Grandchamp:2005yw}
L.~Grandchamp, S.~Lumpkins, D.~Sun, H.~van Hees, and R.~Rapp, ``{Bottomonium
  production at RHIC and CERN LHC},'' {\em Phys.Rev.}, vol.~C73, p.~064906,
  2006.

\bibitem{Emerick:2011xu}
A.~Emerick, X.~Zhao, and R.~Rapp, ``{Bottomonia in the Quark-Gluon Plasma and
  their Production at RHIC and LHC},'' {\em Eur.Phys.J.}, vol.~A48, p.~72,
  2012.

\bibitem{Sharma:2012dy}
R.~Sharma and I.~Vitev, ``{High transverse momentum quarkonium production and
  dissociation in heavy ion collisions},'' 2012.

\bibitem{Oktay:2010tf}
M.~B. Oktay and J.-I. Skullerud, ``{Momentum-dependence of charmonium spectral
  functions from lattice QCD},'' 2010.

\bibitem{Aarts:2012ka}
G.~Aarts, C.~Allton, S.~Kim, M.~P. Lombardo, M.~B. Oktay, {\em et~al.}, ``{S
  wave bottomonium states moving in a quark-gluon plasma from lattice NRQCD},''
  2012.

\bibitem{Nonaka:2011zz}
C.~Nonaka, M.~Asakawa, M.~Kitazawa, and Y.~Kohno, ``{Charmonium spectral
  functions at finite momenta in the gluon plasma from lattice QCD},'' {\em
  J.Phys.}, vol.~G38, p.~124109, 2011.

\bibitem{Aarts:2010ek}
G.~Aarts, S.~Kim, M.~Lombardo, M.~Oktay, S.~Ryan, {\em et~al.}, ``{Bottomonium
  above deconfinement in lattice nonrelativistic QCD},'' {\em Phys.Rev.Lett.},
  vol.~106, p.~061602, 2011.

\bibitem{Aarts:2011sm}
G.~Aarts, C.~Allton, S.~Kim, M.~Lombardo, M.~Oktay, {\em et~al.}, ``{What
  happens to the Upsilon and etab in the quark-gluon plasma? Bottomonium
  spectral functions from lattice QCD},'' {\em JHEP}, vol.~1111, p.~103, 2011.

\bibitem{Song:2007gm}
T.~Song, Y.~Park, S.~H. Lee, and C.-Y. Wong, ``{The Thermal width of heavy
  quarkonia moving in quark gluon plasma},'' {\em Phys.Lett.}, vol.~B659,
  pp.~621--627, 2008.

\bibitem{Asakawa:2000tr}
M.~Asakawa, T.~Hatsuda, and Y.~Nakahara, ``{Maximum entropy analysis of the
  spectral functions in lattice QCD},'' {\em Prog.Part.Nucl.Phys.}, vol.~46,
  pp.~459--508, 2001.

\bibitem{Dominguez:2008be}
F.~Dominguez and B.~Wu, ``{On dissociation of heavy mesons in a hot quark-gluon
  plasma},'' {\em Nucl.Phys.}, vol.~A818, pp.~246--263, 2009.

\bibitem{Zhao:2007hh}
X.~Zhao and R.~Rapp, ``{Transverse Momentum Spectra of J/psi in Heavy-Ion
  Collisions},'' {\em Phys.Lett.}, vol.~B664, pp.~253--257, 2008.

\bibitem{Liu:2006nn}
H.~Liu, K.~Rajagopal, and U.~A. Wiedemann, ``{An AdS/CFT Calculation of
  Screening in a Hot Wind},'' {\em Phys.Rev.Lett.}, vol.~98, p.~182301, 2007.

\bibitem{Chernicoff:2006hi}
M.~Chernicoff, J.~A. Garcia, and A.~Guijosa, ``{The Energy of a Moving
  Quark-Antiquark Pair in an N=4 SYM Plasma},'' {\em JHEP}, vol.~0609, p.~068,
  2006.

\bibitem{Caceres:2006ta}
E.~Caceres, M.~Natsuume, and T.~Okamura, ``{Screening length in plasma
  winds},'' {\em JHEP}, vol.~0610, p.~011, 2006.

\bibitem{Avramis:2006em}
S.~D. Avramis, K.~Sfetsos, and D.~Zoakos, ``{On the velocity and
  chemical-potential dependence of the heavy-quark interaction in N=4 SYM
  plasmas},'' {\em Phys.Rev.}, vol.~D75, p.~025009, 2007.

\bibitem{Natsuume:2007vc}
M.~Natsuume and T.~Okamura, ``{Screening length and the direction of plasma
  winds},'' {\em JHEP}, vol.~0709, p.~039, 2007.

\bibitem{Ejaz:2007hg}
Q.~J. Ejaz, T.~Faulkner, H.~Liu, K.~Rajagopal, and U.~A. Wiedemann, ``{A
  Limiting velocity for quarkonium propagation in a strongly coupled plasma via
  AdS/CFT},'' {\em JHEP}, vol.~0804, p.~089, 2008.

\bibitem{Chernicoff:2012bu}
M.~Chernicoff, D.~Fernandez, D.~Mateos, and D.~Trancanelli, ``{Quarkonium
  dissociation by anisotropy},'' {\em JHEP}, vol.~1301, p.~170, 2013.

\bibitem{CasalderreySolana:2011us}
J.~Casalderrey-Solana, H.~Liu, D.~Mateos, K.~Rajagopal, and U.~A. Wiedemann,
  ``{Gauge/String Duality, Hot QCD and Heavy Ion Collisions},'' 2011.

\bibitem{Myers:2008cj}
R.~C. Myers and A.~Sinha, ``{The Fast life of holographic mesons},'' {\em
  JHEP}, vol.~0806, p.~052, 2008.

\bibitem{Fujita:2009wc}
M.~Fujita, K.~Fukushima, T.~Misumi, and M.~Murata, ``{Finite-temperature
  spectral function of the vector mesons in an AdS/QCD model},'' {\em
  Phys.Rev.}, vol.~D80, p.~035001, 2009.

\bibitem{Carrington:1997sq}
M.~E. Carrington, D.-f. Hou, and M.~H. Thoma, ``{Equilibrium and nonequilibrium
  hard thermal loop resummation in the real time formalism},'' {\em
  Eur.Phys.J.}, vol.~C7, pp.~347--354, 1999.

\bibitem{Bagrinovskii}
B.~K. A. and S.~K. Godunov, ``Difference schemes for multidimensional problems
  (in russian),'' {\em Dokl. Akad. Nauk. USSR}, vol.~115, pp.~431--433, 1957.

\bibitem{crank-nicolson}
J.~{Crank}, P.~{Nicolson}, and D.~R. {Hartree}, ``{A practical method for
  numerical evaluation of solutions of partial differential equations of the
  heat-conduction type},'' {\em Proceedings of the Cambridge Philosophical
  Society}, vol.~43, p.~50, 1947.

\bibitem{harris}
G.~M. {Harris}, ``{Attractive Two-Body Interactions in Partially Ionized
  Plasmas},'' {\em Physical Review}, vol.~125, pp.~1131--1140, Feb. 1962.

\bibitem{rogers}
F.~J. Rogers, H.~C. Graboske, and D.~J. Harwood, ``Bound eigenstates of the
  static screened coulomb potential,'' {\em Phys. Rev. A}, vol.~1,
  pp.~1577--1586, Jun 1970.

\bibitem{Vrscay}
E.~R. {Vrscay}, ``{Hydrogen atom with a Yukawa potential: Perturbation theory
  and continued-fractions-Pad{\'e} approximants at large order},'' {\em \pra},
  vol.~33, pp.~1433--1436, Feb. 1986.

\end{thebibliography}
\end{document}